\pdfoutput=1
\documentclass[fleqn,usenatbib,usedcolumn]{mnras}

\usepackage{graphicx}
\usepackage{multirow}

\newcommand{\teff}{\mbox{$T_{\rm eff}$}}
\newcommand{\logg}{\mbox{$\log g$}}
\newcommand{\vsini}{\mbox{$v \sin i$}}
\newcommand{\mictrb}{\mbox{$\xi_{\rm t}$}}
\newcommand{\mactrb}{\mbox{$v_{\rm mac}$}}

\newcommand{\kms}{\mbox{km\,s$^{-1}$}}
\newcommand{\halpha}{\mbox{$H_\alpha$}}

\title[WASP-92b, WASP-93b and WASP-118b: Three new transiting close-in giant planets]{WASP-92b, WASP-93b and WASP-118b: Three new transiting close-in giant planets}
\author[K. L. Hay et al.]{K. L. Hay$^{1}$\thanks{E-mail: kh97@st-andrews.ac.uk}, A. Collier-Cameron$^{1}$, A. P. Doyle$^{2}$, G. H\'{e}brard$^{3}$, I. Skillen$^{4}$,  
\newauthor D. R. Anderson$^{5}$, S. C. C. Barros$^{6}$, D. J. A. Brown$^{1,2}$, F. Bouchy$^{7}$, R. Busuttil$^{8}$, 
\newauthor P. Delorme$^{1,9}$, L. Delrez$^{10}$, O. Demangeon$^{7}$, R. F. D\'{i}az$^{11}$, M. Gillon$^{10}$, E. Gonz\`{a}lez$^{12}$,
\newauthor C. Hellier$^{5}$, S. Holmes$^{8}$, J. F. Jarvis$^{8}$, E. Jehin$^{7}$, Y. C. Joshi$^{13,14}$, U. Kolb$^{8}$,  
\newauthor M. Lendl$^{11,15}$, P. F. L. Maxted$^{5}$, J. McCormac$^{2}$, G. R. M. Miller$^{1,16}$, A. Mortier$^{1}$, 
\newauthor D. Pollacco$^{2}$, D. Queloz$^{17}$, D. S\'{e}gransan$^{11}$, E. K. Simpson$^{13}$, B. Smalley$^{5}$,
\newauthor J. Southworth$^{5}$, A. H. M. J. Triaud$^{11,17}$, O. D. Turner$^{5}$, S. Udry$^{11}$, M. Vanhuysse$^{18}$,
\newauthor R. G. West$^{2}$, and P. A. Wilson$^{3}$\\
\\
$^{1}$SUPA, School of Physics and Astronomy, University of St Andrews, Fife, KY16 9SS, UK\\
$^{2}$Department of Physics, University of Warwick, Coventry, CV4 7AL, UK\\
$^{3}$Institut d'Astrophysique de Paris, UMR 7095 CNRS, Universit\'{e} Pierre \& Marie Curie, 75000 Paris, France \\
$^{4}$Isaac Newton Group of Telescopes, Apartado de Correos 321, Santa Cruz de La Palma, E-38700, Spain \\
$^{5}$Astrophysics Group, Keele University, Staffordshire, ST5 5BG, UK\\
$^{6}$Instituto de Astrof{\'\i}sica e Ci\^encias do Espa\c{c}o, Universidade do Porto, CAUP, Rua das Estrelas, 4150-762 Porto, Portugal \\
$^{7}$Aix Marseille Universit\'{e}, CNRS, LAM (Laboratoire d'Astrophysique de Marseille) UMR 7326, 13388 Marseille, France \\
$^{8}$Department of Physical Sciences, The Open University, Milton Keynes, MK7 6AA, UK \\
$^{9}$Universit\'{e} Grenoble Alpes, CNRS, IPAG, 38000 Grenoble, France \\
$^{10}$Institut d'Astrophysique et G\'{e}ophysique, Universit\'{e} de Li\`{e}ge, all\'{e}e du 6 Ao\^{u}t 17, B-4000 Li\`{e}ge, Belgium \\
$^{11}$Observatoire de Gen\`{e}ve, Universit\'{e} de Gen\`{e}ve, 51 Chemin des Maillettes, 1290 Sauverny, Switzerland \\
$^{12}$Observatori Astron\'{o}mic de Mallorca, Cam\'{i} de l'Observatori s/n, 07144 Costitx, Mallorca, Spain \\
$^{13}$Astrophysics Research Centre, School of Mathematics \& Physics, Queen's University, Belfast BT7 1NN, UK \\
$^{14}$Aryabhatta Research Institute of Observational Sciences (ARIES), Manora peak, Nainital 263002, India \\
$^{15}$Space Research Institute, Austrian Academy of Sciences, Schmeidlerstr. 6, 8042 Graz, Austria \\
$^{16}$Department of Physics, University of Oxford, Keble Road, Oxford, OX1 3RH, UK \\
$^{17}$Institute of Astronomy, University of Cambridge, Madingley Road, Cambridge, CB3 0HA, United Kingdom \\
$^{18}$OverSky, 47 all\'{e}e des Palanques, BP 12, 33127, Saint-Jean d'Illac, France \\
}

\begin{document}

\date{Accepted year month day. Received year month day; in original form year month day}

\pagerange{\pageref{firstpage}--\pageref{lastpage}} \pubyear{2016}

\maketitle

\label{firstpage}

\begin{abstract}
We present the discovery of three new transiting giant planets, first detected with the WASP telescopes, and establish their planetary nature with follow up spectroscopy and ground-based photometric lightcurves. WASP-92 is an F7 star, with a moderately inflated planet orbiting with a period of 2.17 days, which has $R_p = 1.461 \pm 0.077 R_{\rm J}$ and $M_p = 0.805 \pm 0.068 M_{\rm J}$. WASP-93b orbits its F4 host star every 2.73 days and has $R_p = 1.597 \pm 0.077 R_{\rm J}$ and $M_p = 1.47 \pm 0.029 M_{\rm J}$. WASP-118b also has a hot host star (F6) and is moderately inflated, where $R_p = 1.440 \pm 0.036 R_{\rm J}$ and $M_p = 0.513 \pm 0.041 M_{\rm J}$ and the planet has an orbital period of 4.05 days. They are bright targets (V = 13.18, 10.97 and 11.07 respectively) ideal for further characterisation work, particularly WASP-118b, which is being observed by K2 as part of campaign 8. WASP-93b is expected to be tidally migrating outwards, which is divergent from the tidal behaviour of the majority of hot Jupiters discovered.

\end{abstract}


\section{Introduction}
The WASP consortium \citep{pollacco2006} has been highly successful at identifying and confirming the existence of transiting hot Jupiters (orbital period, P$<$10 days). Wide angle ground-based surveys are able to find rarer objects with small orbital separations and deep transits found predominantly around stars brighter than V=13. This wealth of discoveries of bright stars hosting planets has enabled detailed follow up observations. The planets discovered by these surveys dominate the targets used for planetary atmosphere characterisation and further understanding of planet formation mechanisms and migration e.g. \citet{sing2016,ford2006,matsumura2010}.

Hot Jupiters are rare objects with an occurrence rate estimated to be $\sim1\%$, depending on the stellar population surveyed \citep{wang2015,howard2012,wright2012}. Searches for close in companions to hot Jupiters via transit timing variations (TTV; \citet{steffen2012}) have been unsuccessful. This lack of close in planets points towards the theory that these objects form outside the snow line and migrate towards the orbits they are observed in through high eccentricity migration (HEM) \citep{rasio1996,fabrycky2007,mustill2015}. 

It was first noted by \citet{winn2010} and expanded upon by \citet{albrecht2012} that the distribution of projected hot Jupiter orbital obliquities appeared correlated with the temperature of the host star. The two groups of obliquities seen were separated at a stellar effective temperature around $6250$ K, with cooler stars more likely to host hot Jupiters with orbits aligned to the stellar rotation axis, and hotter stars with a range of alignments. The range of alignments observed is another indicator that HEM is a strong candidate for the prevailing migration mechanism for hot Jupiters with strong misalignments.

Recent statistical work has however shown that not all hot Jupiters can have undergone HEM due to the lack of super-eccentric orbits found \citep{dawson2015}. The K2 mission \citep{howell2014} allows for the detection of the transits of small companion planets for hot Jupiters. This capability was demonstrated with the discovery of 2 further planets in close in orbits around WASP-47 \citep{becker2015}, which leads to the question of whether close in planets can be detected with space-based photometry for other systems with hot Jupiters. 


In this paper we present the discovery of WASP-92b, WASP-93b and WASP-118b. The latter system will be observed in campaign 8 of the K2 mission. All three of these planetary systems are excellent candidates for spin-orbit alignment follow up.

Section \ref{sec:obs} introduces the observational data collected for the systems in this paper. Section \ref{sec:stellar} presents the spectral analyses of the host stars, and Section \ref{sec:analy} describes the methods used to determine the parameters of the newly discovered systems. The discoveries and their tidal evolution are discussed in Section \ref{sec:discu}.

\section{Observations}
\label{sec:obs}
\subsection{WASP photometery}

The newly discovered planets presented in this paper were all initially detected from observations of the WASP telescopes -- the northern installation (SuperWASP) for WASP-92 and 93, and jointly between north and south (WASP-S) for WASP-118.

WASP-92 (1SWASPJ162646.08+510228.2; 
V=13.18; K=11.52) was observed with SuperWASP from 2007-03-30 to 2010-08-03. 
WASP-93 (1SWASPJ003750.11+511719.5; 
V=10.97; K=9.94) was observed with SuperWASP from 2007-07-26 to 2007-12-25. 
WASP-118 (1SWASPJ011812.12+024210.2; 
V=11.02; K=9.79) was observed on both SuperWASP and WASP-S between 2008-07-13 and 2010-12-13. 
The seasonal intervals and quantities of photometric points for these measurements are shown in Table \ref{tab:wasp_phot}.

Photometry for these systems was detrended and searched for transit signals using the algorithms described in \citet{cameron2006}. All three were selected as high priority candidates for follow-up photometry and spectroscopy using the treatment of \citet{cameron2007}. The key component of the candidate identification is to search the photometric data collected by the WASP telescopes for a periodic transit signal using an adaptation of the \citet{kovacs2002} Box-Least Squares (BLS) algorithm. WASP-92 was selected with a 2.175 day periodic transit signature; WASP-93 with a 2.732 day period, and WASP-118 with a 4.046 day period. The upper plots in Figures \ref{fig:wasp92} -- \ref{fig:wasp118} show the WASP photometry folded onto the best-fit orbital ephemeris for each system, found in the global system analysis in Section \ref{sec:analy}. Due to variation introduced by the observational systematic errors, photon noise and from stellar activity, the scatter in the photometric time series data is fairly large, thus the WASP photometry is also plotted in 200 phase bins. Each bin corresponds to $1/200^{\textrm{th}}$ of the orbital period, so that the transit shape observed can be clearly seen on the plot.

\begin{table}
\centering
\caption{Details of the observing intervals of the systems with the WASP telescopes}
\begin{tabular}{ccc}
\hline
\hline
System & Interval & N$_{points}$ \\
\hline
WASP-92 & 2007-03-30 -- 2007-08-04 & 11674 \\
& 2008-03-28 -- 2008-08-03 & 8551 \\
& 2009-03-31 -- 2009-08-03 & 10358 \\
& 2010-03-28 -- 2010-08-03 & 11693 \\
\hline
WASP-93 & 2007-07-26 -- 2007-12-25 & 8614 \\
\hline
WASP-118 & 2008-07-13 -- 2008-12-13 & 7750 \\
& 2009-07-17 -- 2009-12-09 & 9297 \\
& 2010-08-18 -- 2010-12-13 & 4906 \\
\hline
\hline
\end{tabular}
\label{tab:wasp_phot}
\end{table}

\subsection{Photometric follow-up}
Transits of the systems were also observed with a number of different telescopes at a higher cadence than with the WASP cameras to better determine the transit shapes. The dates and details for these observations are shown in Table \ref{tab:phot_dates}.

\begin{table*}
\begin{centering}
\caption{Details of photometric follow up observations with telescope/instrument used, and the date at start of observation}
\begin{tabular}{cccccccc} 
\hline
\hline
System & Observation start & Instrument & Filter & Exposure time (s) & N$_{\textrm{points}}$ & Defocus? & Full transit? \\ \hline
WASP-92 & 2012-07-24 & OverSky & R band & 90 & 179 & N & N \\
& 2013-06-26 & NITES & none & 20 & 751 & Y & Y \\
& 2013-07-09 & RISE & V+R band & 20 & 743 & Y & Y \\
& 2013-07-22 & NITES & none & 20 & 630 & Y & Y \\
\hline
WASP-93 & 2008-07-10 & RATCam & Z band & 10 & 144 & Y & Y \\
& 2009-01-23 & JGT & R band & 60 & 156 & N & N \\
& 2011-09-28 & JGT & R band & 15 & 627 & N & N \\
& 2011-10-31 & PIRATE & R band & 45 & 142 & N & Y \\
& 2012-01-21 & JGT & R band & 30 & 458 & N & Y \\
& 2012-07-22 & PIRATE & R band & 120 & 87 & N & Y \\
& 2012-08-21 & NITES & none & 15 & 785 & Y & Y \\
& 2013-10-05 & RISE & V+R band & 3.5 & 3717 & Y & Y \\
\hline
WASP-118 & 2013-10-27 & EulerCam & z' Gunn & 60 & 226 & Y & Y \\
& 2014-10-11 & TRAPPIST & I+z band & 7 & 1029 & N & N \\
& 2014-10-18 & EulerCam & z' Gunn & 50 & 264 & Y & Y \\
& 2015-10-01 & EulerCam & I band & 60 & 215 & Y & N \\
\hline
\hline 
\\
\end{tabular}
\label{tab:phot_dates}
\newline {\bf Note:} Details of the filters used for each observation are described in more detail with the descriptions of observing procedures for each telescope, where values for amount of defocussing used are also noted. Transit coverage can be seen in the lightcurves for each observation in the second panels of Figures \ref{fig:wasp92}--\ref{fig:wasp118}
\end{centering}
\end{table*}

Individual telescopes and instruments used for the transit observations have their own observing techniques and data reduction pipelines, which are briefly described in the sections below. The reduced lightcurves of these observations for each system are shown in the second plots in Figures \ref{fig:wasp92} -- \ref{fig:wasp118}, where all of the data are phased with reference to the epoch of mid-transit.

\subsubsection{Oversky observations}
\emph{OverSky} is a 0.36 m robotic telescope in La Palma. It was used for a partial transit of WASP-92b on the night beginning 2012 July 24. The telescope is installed with an SBIG STL-1001E, which contains a $1024\times1024$ pixels ($24\mu$) Kodak KAF-1001E CCD, with active optic SBIG AOL, and using internal guiding during observations. The telescope was in focus throughout the observations, and 179 images of the $19.9\times19.9$ arcmin field of view (FOV) were take with a Sloan $r'$ filter. Each image was the result of a 90 second exposure with 7 seconds dead time between each for CCD readout.

The images were bias, dark and dusk flat calibrated, and the lightcurve was extracted from the images using Munipack Muniwin 2.0, using 2 reference stars and 1 check star in the field.

\subsubsection{NITES observations}
\emph{NITES} (\emph{N}ear \emph{I}nfra-red \emph{T}ransiting \emph{E}xoplanet\emph{S}) is a semi-robotic $0.4$-m (f/10) Meade LX200GPS Schmidt-Cassegrain telescope installed at the ORM, La Palma. The telescope is mounted with Finger Lakes Instrumentation Proline 4710 camera, containing a $1024\times1024$ pixels deep-depleted CCD made by e2v. The telescope has a FOV and pixel scale of $11\times11$ arcmin squared and $0.66$ arcsec pixel$^{-1}$, respectively and a peak quantum efficiency (QE) $>90\%$ at $800$ nm \citep{mccormac2014}. The observations were all made without a filter, and with the telescope defocussed to $3.3$ arcsec FWHM for the transit of WASP-92 on 2013 June 26; $2.8$ arcsec FWHM on 2013 July 22, and to $6.6$ arcsec for the transit of WASP-93 on 2012 August 21.

The data were bias subtracted and flat field corrected using PyRAF\footnotemark \footnotetext{PyRAF is a product of the Space Telescope Science Institute, which is operated by AURA for NASA.} and the standard routines in IRAF\footnotemark \footnotetext{IRAF is distributed by the National Optical Astronomy Observatories, which are operated by the Association of Universities for Research in Astronomy, Inc., under cooperative agreement with the National Science Foundation.} and aperture photometry was performed using DAOPHOT \citep{stetson1987}. For WASP-92b $8$ nearby comparison stars were used, and $5$ for WASP-93b. Aperture radii of $5.9$ arcsec, $5.0$ arcsec and $10.2$ arcsec were chosen for 2013 June 26, 2013 July 22 and 2012 August 21 respectively. Aperture sizes were chosen to minimise the RMS scatter in the out of transit data. Initial photometric error estimates were calculated using the electron noise from the target and the sky and the readout noise within the aperture. The data were normalised with a first order polynomial fitted to the out-of-transit data. 

\subsubsection{RISE observations}
\emph{RISE} (\emph{R}apid \emph{I}maging \emph{S}earch for \emph{E}xoplanets) is an optical camera on the 2-m robotic Liverpool Telescope in La Palma \citep{steele2008,gibson2008}. The camera set up consists of a $1024\times1024$ pixel e2v CCD, which has a $9.4\times9.4$ arcmin FOV, and a single fixed filter that equates to V+R band (500--700 nm). The images were debiased and flat-fielded using twilight flats using IRAF routines. The photometric lightcurves were then extracted for the target and a small number of nearby non-variable comparison stars, and differential photometry performed, all utilising PyRAF and DAOPHOT routines.

\subsubsection{RATCam observations}
\emph{RATCam} is an optical CCD imager, which was installed on the Liverpool Telescope until decomissioned in February 2014 \citep{steele2004,mottram2004}. The camera was a $2048\times2048$ pixel EEV CCD, set up with a $4.6\times4.6$ arcmin FOV. 

For the WASP-93 lightcurve observed on the night beginning 2008-07-10, images were taken in 10 s exposures, which were bias subtracted and flat-fielded using a recent twilight flat-field image. Aperture photometry was performed using IRAF DAOPHOT routines with three bright non-variable comparison stars in the observed field. The lightcurve was then binned into sets of three images, resulting in 144 photometric data points.

\subsubsection{PIRATE observations}
PIRATE (Physics Innovations Robotic Astronomical Telescope Explorer) is a robotic 0.43-m (f/6.8) corrected Dall-Kirkham telescope, located at the Observatori Astronomic de Mallorca at the time of the observations \citep{holmes2011,kolb2014}. The facility has a choice of two imaging cameras, the SBIG STL-1001E, which contains a $1024 \times 1024$ pixel Kodak KAF-1001E CCD, and the SBIG STX-16803, which contains a $4096 \times 4096$ pixel Kodak KAF-16803 CCD. These two CCDs describe PIRATE's two operating modes, Mk 1.5 and Mk 2 respectively. The Mk 2 FOV is $43 \times 43$ arcmin squared with a pixel scale of 0.63 arcsec pixel$^{-1}$ and the peak QE is 0.60 at 550 nm.

The transit observations using PIRATE were made using the Mk 2 operating mode with a Baader R broadband filter which is equivalent to the APASS r filter, and without a telescope defocus. The 2011-10-31 data set was obtained in 2x2 pixel binning while the rest of the data sets were obtained without binning. The images were bias, dark and dusk flat calibrated, before ensemble photometry was performed using DAOPHOT as described in \citet{holmes2011}.

\subsubsection{JGT observations}
JGT (James Gregory Telescope) is the 0.94-m telescope at the University of St Andrews Observatory. The CCD is an 1024x1024 e2v device, and the telescope has an un-vignetted FOV of $15$ arcmin diameter. The JGT data was collected using a standard R-band filter and the telescope in focus. The data collected were corrected with flat-field and bias images, and then relative aperture photometry with a single comparison star was performed using the GAIA routines of the Starlink software packages\footnotemark \footnotetext{Starlink is maintained by the East Asian Observatory, and is published as open source at http://starlink.eao.hawaii.edu/starlink}.

\subsubsection{EulerCam observations}
\emph{EulerCam} is the camera on the 1.2-m Swiss-Euler Telescope in La Silla, Chile \citep{lendl2012}. The CCD is a 4kx4k e2v deep-depletion silicon device, used in a set up which produces images with a FOV of $15.68 \times 15.73$ arcmin. Flat-field uncertainties can be a major contributor to high-precision photometry uncertainties, so a precise tracking system is deployed from exposure to exposure to ensure the stellar images remain on the same CCD pixels.

The images were overscan, bias and flat-field corrected before ensemble photometry was performed with 3--10 reference stars. The number of reference stars similar in colour and magnitude to the target used was decided with an iterative process to minimise RMS scatter in the transit lightcurve.

\subsubsection{TRAPPIST observations}
\emph{TRAPPIST} (\emph{TRA}nsiting \emph{P}lanets and \emph{P}lanetes\emph{I}mals \emph{S}mall \emph{T}elescope) is the 0.6 m robotic telescope in La Silla, Chile operated by Li\`{e}ge University \citep{jehin2011}. The installed CCD is 2048x2048 pixels, with a FOV of $22 \times 22$ arcmin. The I+z filter used for the TRAPPIST observations has $>$90\% transmission from 750 to 1100 nm, with the long wavelength cutoff being set by the QE of the CCD detector.

During the use of the telescope, astrometric solutions of the images are used to send pointing corrections to the mount to retain the stellar images on the same pixels throughout observations. After bias, dark and flat-field correction, aperture photometry was extracted from the images using IRAF/DAOPHOT2 routines \citep{stetson1987}. After a careful selection of reference stars similar in brightness to the target star, lightcurves were obtained using differential photometry.

\subsection{Radial velocity follow-up}
\label{sec:spec}

\begin{table}
\centering
\caption{Details of the observing intervals of the RV follow up}
\begin{tabular}{cccc}
\hline
\hline
System & Interval & Instrument &N$_{points}$ \\
\hline
WASP-92 & 2012-05-23 -- 2012-07-26 & SOPHIE & 9 \\
\hline
WASP-93 & 2008-07-15 -- 2010-11-28 & SOPHIE & 13 \\
& 2010-08-17 -- 2010-08-18 & SOPHIE & 34 \\
& 2011-08-18 -- 2011-08-19 & SOPHIE & 27 \\
\hline
WASP-118 & 2010-10-16 -- 2010-10-19 & SOPHIE & 4 \\
& 2010-07-19 -- 2014-08-26 & CORALIE & 23 \\
& 2011-07-29 -- 2011-07-31 & SOPHIE & 2 \\
& 2015-07-23 -- 2015-11-07 & CORALIE & 4 \\
& 2015-10-01 -- 2015-10-02 & HARPS-N & 25$^{\star}$ \\
& 2015-10-04 -- 2015-10-06 & HARPS-N & 3\\
\hline
\hline
\end{tabular}
\label{tab:rv_intervals}
\newline {\bf Note:} Observations have been grouped according to the instrumental configurations used. $^{\star}$Data set not included in the global fit for the determination of the final parameters presented.
\end{table}

In order to measure the radial velocity (RV) reflex motion on the star from the planet's orbit, each of the host stars were observed on multiple occasions with the SOPHIE spectrograph \citep{perruchot2008,bouchy2009}. The SOPHIE spectrograph is installed on the 1.93-m telescope of the Observatoire de Haute-Provence, France. The second fibre is pointed towards the sky to counteract the effect of the moon and sky. SOPHIE has two observing modes, high resolution and high efficiency, so data collected in each mode are considered to be from different instruments that could have a baseline RV offset from each other \citep{bouchy2009}.

WASP-118 was also observed with the CORALIE spectrograph \citep{baranne1996,queloz2000}, which is mounted on the Swiss Euler 1.2-m telescope. CORALIE points its second spectrograph fibre on a calibration lamp, and lunar contamination is avoided by observing only when the moon is not present. The CORALIE set-up was upgraded in November 2014, thus data from before and after this upgrade are treated as separate instruments, which could have an offset in the measured RVs. 

For data from each of the spectrographs, radial velocities were computed from the spectra by weighted cross-correlation (CCF) \citep{pepe2002} using a numerical G2-spectral template.

WASP-118 has been further observed with the HARPS-N instrument on the 3.6-m Telescopio Nazionale Galileo (TNG) on La Palma \citep{cosentino2012}. An attempt was made to collect spectra during a transit on 1st October 2015 of the object to determine orbital obliquity, but cloud cover prevented data collection. Three further spectral observations were made during the following orbit, two of which were during the transit. As with SOPHIE observations, the second spectrograph fibre was fed from an off-target aperture in the telescope focal plane to allow accurate sky background subtraction. These observations were reduced with the data reduction software (DRS) version 3.7, for which one of the outputs is RV measurements calculated from a Gaussian fit to the mean CCF obtained from each spectral order.

\begin{figure}
\includegraphics[width=0.48\textwidth]{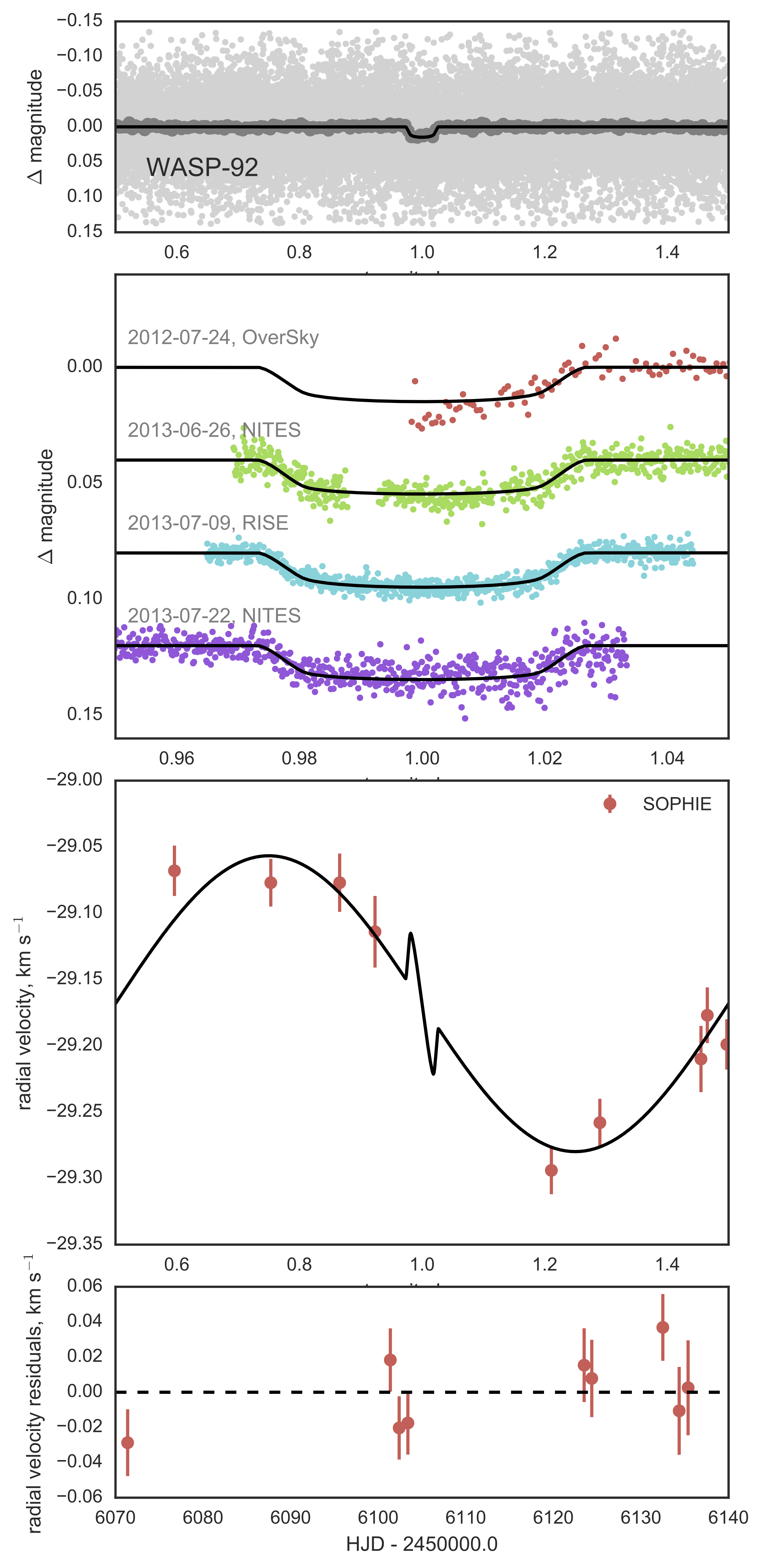}
\caption{Figure showing the observations collected for WASP-92 and used in the global fit described in Section \ref{sec:analy}. The upper plot is of the photometry in collected by the WASP cameras in light grey, which is folded onto best-fit orbital ephemeris presented in this work. The folded data is overlaid in dark grey with the data binned in 1/200th of the phase for clarity in the transit shape, and the transit model is shown in black. The upper middle plot shows the follow up photometry with reference to transit phase, the order presented is the same as for Table \ref{tab:phot_dates}. The lower middle plot shows the RV points with reference to transit phase, and the residuals of the fit with reference to time are shown in the lower plot.}
\label{fig:wasp92}
\end{figure}

\begin{figure}
\includegraphics[width=0.48\textwidth]{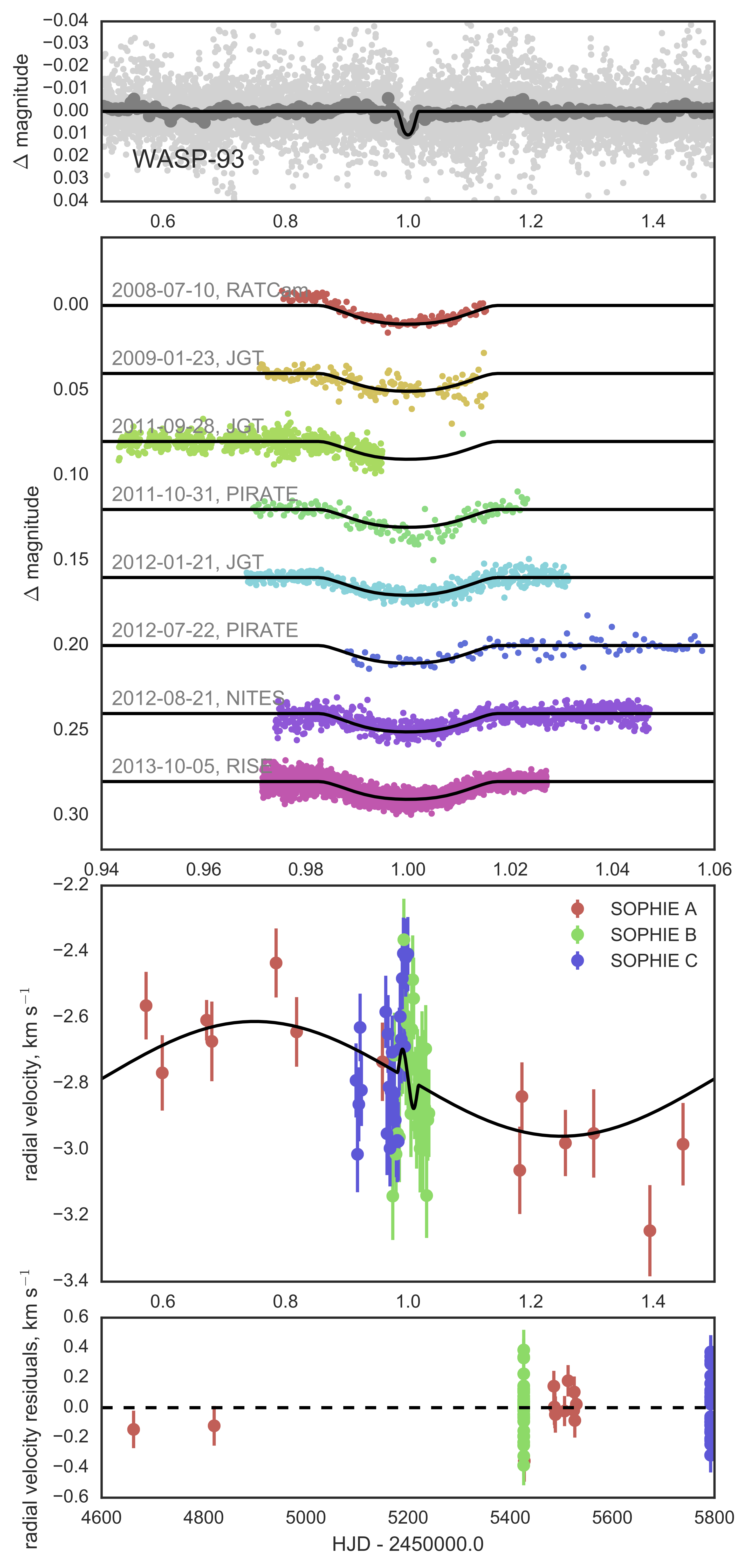}
\caption{Figure showing the data collected for the WASP-93 system. The plots are presented in the same way as for WASP-92 in Figure \ref{fig:wasp92}.}
\label{fig:wasp93}
\end{figure}

\begin{figure}
\includegraphics[width=0.48\textwidth]{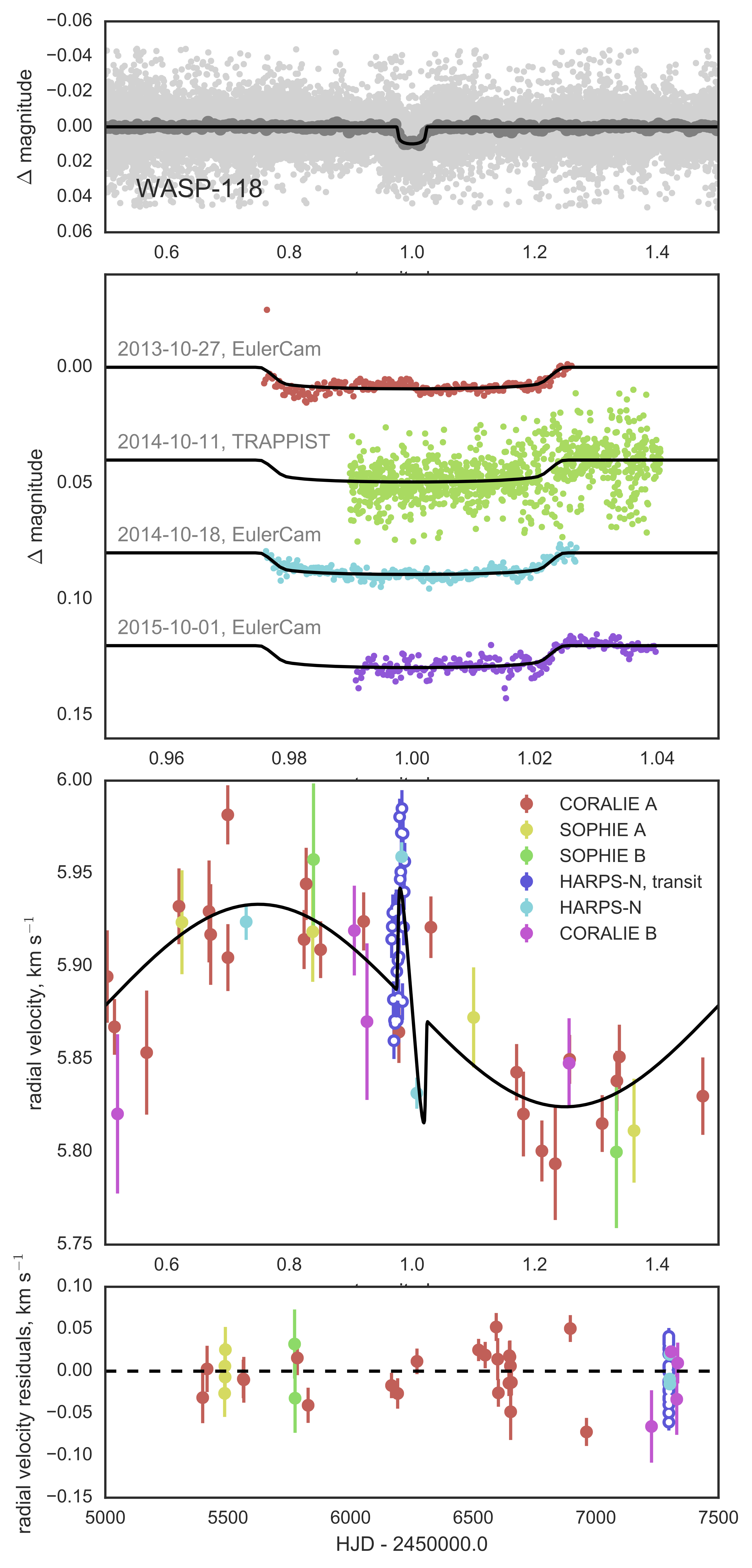}
\caption{Figure showing the observations made of WASP-118, which are presented in the same way as described in the caption of Figure \ref{fig:wasp92}. The HARPS-N data taken as a series during a transit (shown in purple) have open circle markers to denote that the observations were not included in the global fit.}
\label{fig:wasp118}
\end{figure}

The lower two plots in Figures \ref{fig:wasp92} -- \ref{fig:wasp118} show the RV data collected for these systems plotted on the best fit model found in this work, and the residuals of this fit. The third plot in Figure \ref{fig:wasp93} shows a clustering of the RV data around the transit section of the orbital phase, which is due to two attempts to observe the Rossiter-McLaughlin (RM) effect \citep{rossiter1924,mclaughlin1924}.

From the RV measurement determination, measurements for different stellar activity indicators were extracted from the CCFs: full width at half maximum (FWHM), line bisector inverse slope (BIS) and contrast. The uncertainties have been scaled from the RV error as presented in \citet{santerne2015}. These measurements were not possible using the spectra taken with SOPHIE of WASP-93 due to the very broad CCF resulting from the fast rotation of the star.

\subsection{Adaptive optics imaging follow up}

\begin{figure}
\includegraphics[width=0.45\textwidth]{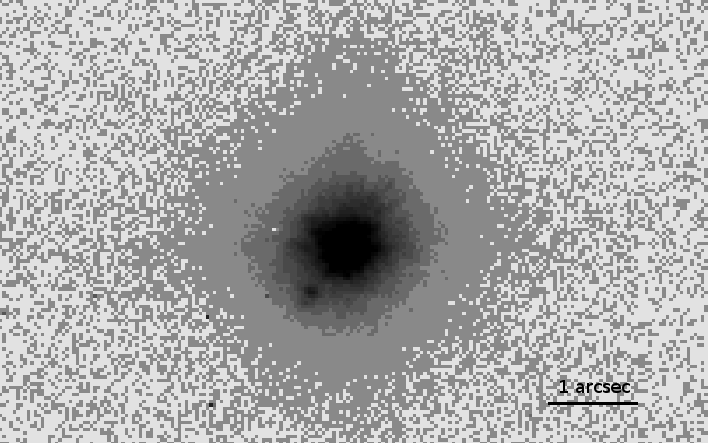}
\caption{H-band image of WASP-93 with correction to 0.14 arcsec, which was taken with INGRID on the WHT in 2011. This image was observed in telescope sky position angle 0 $^{\circ}$ North is up, and East is to the right. The position of the faint companion relative to North remained unchanged in images at different sky position angles, confirming that this companion is not an artefact of the PSF.}
\label{fig:ao_wasp93}
\end{figure}

To investigate blend scenarios we acquired high-resolution J, H and Ks band images of WASP-93 on the night of 2011, August 13 with the near-infrared camera INGRID fed by the NAOMI adaptive-optics system on the 4.2-m William Herschel Telescope. Images were taken in multiple sky position angles (0, 20 and 35 degrees) in order to rule out false positive detections related to PSF features. Natural (open loop) seeing ranged from 0.6 to 0.8 arcsec in the H band, and images with corrected FWHM 0.15 arcsec in the H and Ks bands, and 0.20 arcsec in the J band, were secured. The H-band image is shown in Figure \ref{fig:ao_wasp93}.

These images show that WASP-93 has a nearby stellar companion 0.69$\pm$0.01 arcsec away in position angle 220.89+/-0.60 degrees, which is fainter by 3.70$\pm$0.18, 3.45$\pm$0.10 and 3.37$\pm$0.13 magnitudes respectively in the J, H and Ks bands. Assuming the 2MASS magnitudes of WASP93, this companion has J=13.93$\pm$0.18, H=13.45$\pm$0.10 and Ks=13.31$\pm$0.13, and if it is associated with WASP-93 is an early-to-mid K spectral type dwarf star.

No other sources are detected in the 40 arcsec field-of-view of INGRID to 5-sigma limiting magnitudes of J $\sim$ 19.0$\pm$0.35, H$\sim$19.0$\pm$0.35 and Ks$\sim$18.2$\pm$0.35 at distances $>$ $4\times$FWHM from the centre of the corrected profile, or to $\sim$2 magnitudes brighter than these limits down to $1.5\times$FWHM from the stellar core.


\section{Stellar parameters from spectra}
\label{sec:stellar}

\subsection{WASP-92 stellar parameters}
\label{sec:stellar92}

Individual SOPHIE spectra with no moon pollution of WASP-92 were co-added to produce a single spectrum with a typical signal to noise ratio (SNR) of around 25:1. The standard pipeline reduction products were used in the analysis. The analysis was performed using the methods given in \citet{doyle2013}. The parameters obtained from the analysis are listed in the spectroscopic section of Table~\ref{tab:parameters}. The effective temperature (\teff) was determined from the excitation balance of the Fe~{\sc i} lines. The Na~{\sc i} D lines and the ionisation balance of Fe~{\sc i} and Fe~{\sc ii} were used as surface gravity (\logg) diagnostics. The iron abundance was determined from equivalent width measurements of several unblended iron lines. A value for microturbulence (\mictrb) was determined from Fe~{\sc i} using the method of \cite{magain1984}. The quoted error estimates include that given by the uncertainties in \teff\ and \logg, as well as the scatter due to measurement and atomic data uncertainties. Interstellar Na D lines are present in the spectra with equivalent widths of $\sim$0.02\AA, indicating an extinction of $E(B-V)$ = 0.005 $\pm$ 0.001 using the calibration of \cite{munari1997}.

The projected stellar rotation velocity (\vsini) was determined by fitting the profiles of several unblended Fe~{\sc i} lines. A value for macroturbulence (\mactrb) of 4.97 $\pm$ 0.73 {\kms} was determined from the calibration of \citet{doyle2014}. An instrumental FWHM of 0.15 $\pm$ 0.01~{\AA} was determined from the telluric lines around 6300\AA. A best fitting value of \vsini\ = 5.73 $\pm$ 1.15 ~\kms\ was obtained. 

Lithium is detected in the spectra, with an equivalent width upper limit of 52m\AA, corresponding to an abundance upper limit of $\log A$(Li) 2.70 $\pm$ 0.09. This implies an age of several~Myr \citep{sestito2005}.

The rotation rate ($P = 10.07 \pm 2.58$~d) implied by the {\vsini} and stellar radius gives a gyrochronological age of $\sim$2.29$^{+6.80}_{-1.51}$~Gyr using the \citet{barnes2007} relation. 

\subsection{WASP-93 stellar parameters}
\label{sec:stellar93}

The analysis for WASP-93 was performed as for WASP-92 with co-added spectra with no moon pollution. This produced a master spectrum with a typical SNR around 50:1, however the high \vsini\ of WASP-93 required some different techniques to determine stellar parameters. For example, the \vsini\ was too high to measure a sufficient number of Fe lines in order to obtain the effective temperature (\teff) from the excitation balance. As the SOPHIE spectra were unsuitable for measuring the Balmer lines, an ISIS spectrum of the \halpha\ region was used instead. The high stellar rotation also meant that determining a value for microturbulence (\mictrb) was not possible. For WASP-93, the interstellar Na D lines indicate an extinction of $E(B-V)$ = 0.07 $\pm$ 0.02.

It should also be noted that whilst a value for macroturbulence (\mactrb) was found using the same method as above to be 6.95 $\pm$ 0.73 {\kms}, it has a negligible influence on line broadening with such a high \vsini. The same instrumental FWHM was found in the WASP-93 data.

Unlike WASP-92, there is no significant detection of lithium in the spectra, with an equivalent width upper limit of 0.01 m\AA, corresponding to an abundance upper limit of $\log A$(Li) $<$ 1.14 $\pm$ 0.09. This implies an age of several~Gyr \citep{sestito2005}.

The rotation rate ($P = 1.45 \pm 0.4$~d) implied by the {\vsini} and stellar radius gives a gyrochronological age of $\sim$0.70 $\pm$ 0.65~Gyr using the \citet{barnes2007} relation.

\subsection{WASP-118 stellar parameters}
\label{sec:stellar118}

The co-added CORALIE spectrum of WASP-118 was also analysed using the same techniques as for WASP-92 with no moon pollution. The typical SNR of the spectrum was around 50:1. For WASP-92, the interstellar Na D lines indicated an extinction of $E(B-V)$ = 0.10 $\pm$ 0.03, using the same technique as above. A value for macroturbulence (\mactrb) of 5.77 $\pm$ 0.73 {\kms} was determined, and the instrumental FWHM of 0.11 $\pm$ 0.01~{\AA} was found from the telluric lines around 6300\AA.

There is no significant detection of lithium in the spectra of WASP-118, with an equivalent width upper limit of 4m\AA, corresponding to an abundance upper limit of $\log A$(Li) $<$ 1.21 $\pm$ 0.09. This implies an age of several~Gyr \citep{sestito2005}.

The rotation rate ($P = 6.12 \pm 1.11$~d) implied by the {\vsini} and stellar radius gives a gyrochronological age of $\sim$1.17$^{+5.72}_{-0.75}$~Gyr using the \citet{barnes2007} relation.

\subsection{Stellar activity analysis}

Each of the WASP photometric data series was searched for periodic modulations, which could be attributed to stellar rotation, using the sine-wave fitting method described in \citet{maxted2011}. No signals down to amplitudes of 3 mmag were detected in the lightcurves of WASP-92, 93 and 118, which indicates that stellar spot activity on these three planet-hosts is at a low level.

\begin{figure}
\includegraphics[width=0.48\textwidth]{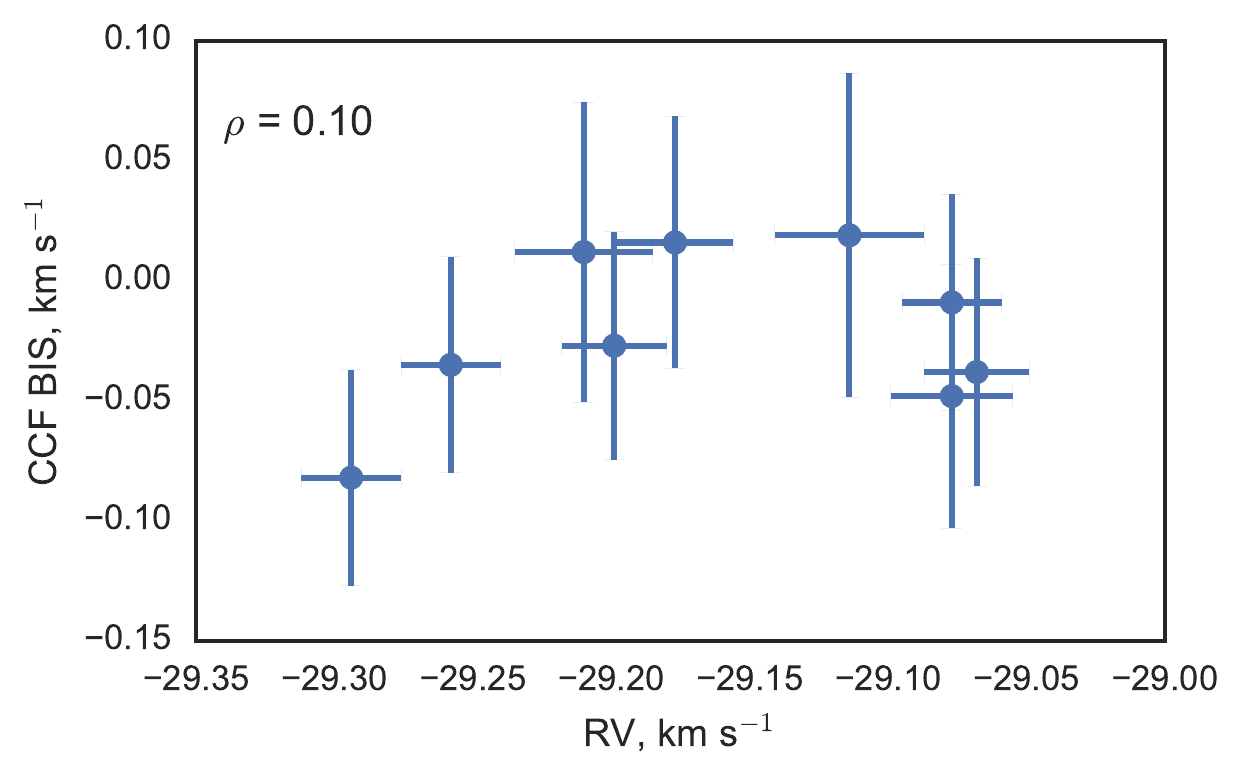}
\includegraphics[width=0.48\textwidth]{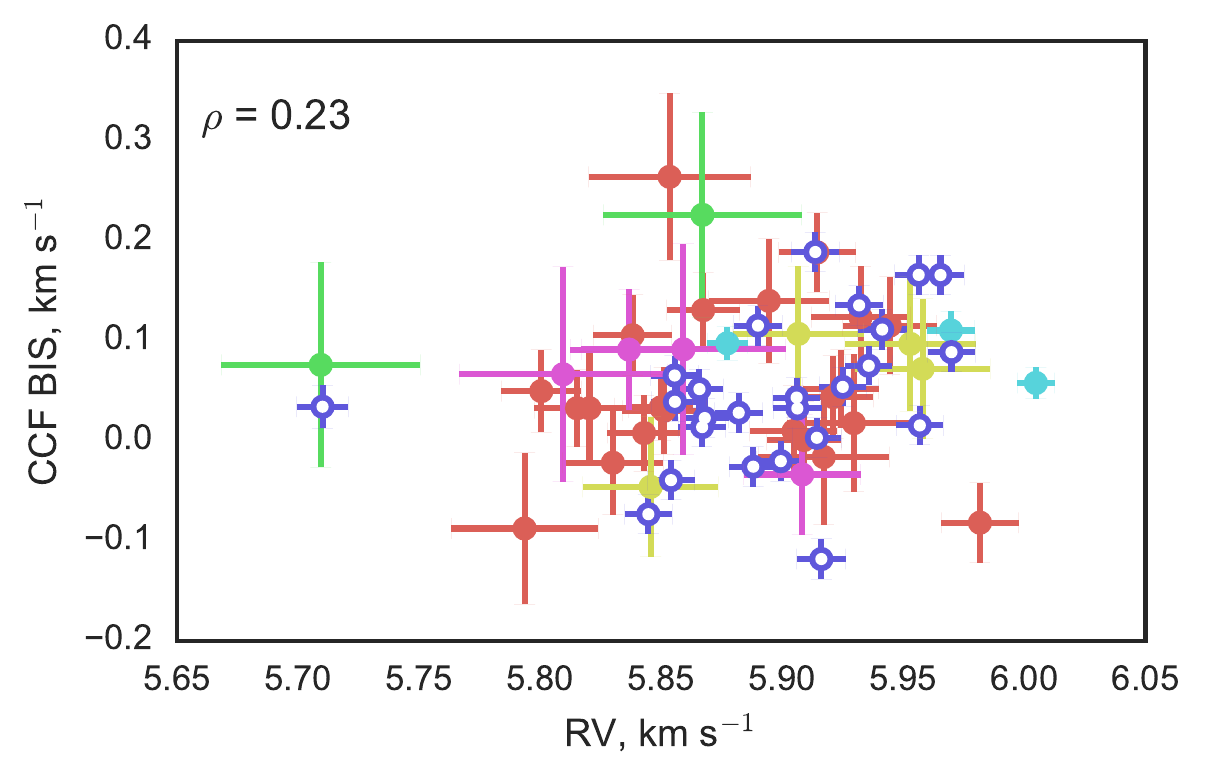}
\caption{The distribution of CCF BIS with measured RV for WASP-92 (upper plot) and WASP-118 (lower plot), where the colours are equivalent to those used for the RV data in Figures \ref{fig:wasp92} and \ref{fig:wasp118}. The Spearman's rank correlation coefficients ($\rho$) are shown in the upper left of the plots.}
\label{fig:bisector}
\end{figure}

Whilst the stability in time of the transit signatures in the photometry indicates that stellar spot transits are not being mistaken for planets, it is useful to also investigate the stellar activity indicators in the collected spectra. A correlation between RV and BIS would be an indicator that the periodicity seen in the RV is caused by stellar activity, or a blended background binary. BIS measurements were not available for WASP-93, as discussed in \ref{sec:spec}. Figure \ref{fig:bisector} shows the distribution of RV against BIS, where low correlation is seen between these quantities. The Spearman's rank correlation coefficients for each are below 0.25, which confirms our confidence that the observed periodic RV signal is of a planetary nature.

\section{Planetary system parameter determination}
\label{sec:analy}

The majority of the orbital system parameter fitting performed for planets discovered by the WASP consortium uses a Markov Chain Monte Carlo (MCMC) algorithm based on the method described by \citet{cameron2007}, and expanded upon in \citet{pollacco2008}. The key component of this algorithm is that the values of the selected parameters perform a random walk in the parameter space, which allows the joint posterior probability distribution of the fitting parameters to be mapped. 

The transit lightcurves are modelled using the analytic formulae of \citet{mandel2002}, which calculates the amount of light obscured by two limb-darkened overlapping discs. In this case the formulation was used with the non-linear limb darkening law with the coefficients determined from the tables of \citet{claret2000} at each step depending on the stellar temperature sampled. The RV fit is performed by fitting a Keplerian to the RV data points, and the RM effect (when observations are available) is modelled with the equations presented by \citet{ohta2005}. Each individual RV dataset is assigned a floating instrumental zero-point velocity of its own, which is determined at each MCMC trial by inverse variance weighted averaging of the RV residuals following subtraction of the orbital velocities derived from the current model - for the fits presented here, the zero-point differences calculated between instruments were all smaller than $0.1$ m s$^{-1}$.
 
The jump parameters used in the MCMC analysis are listed with their definitions in Table \ref{tab:jump}. The parameters used have been updated from those listed in \citet{cameron2007} by \citet{enoch2010} and \citet{anderson2011} to ensure that truly uniform priors are used. Where an eccentric orbital solution is allowed, $\sqrt{e}\cos\omega$ and $\sqrt{e}\sin\omega$ were also used as jump parameters, where $e$ is orbital eccentricity and $\omega$ is the argument of periastron. With modelling of the RM effect, $\sqrt{v \sin i}\sin \lambda$ and $\sqrt{v \sin i}\cos \lambda$ were used as jump parameters, where $\lambda$ is the projected angle between the orbital motion of the planet and the rotational motion of the star. The particular formulation of jump parameters has been selected to ensure as little correlation between parameters as possible, and so that each jump is in a uniform prior probability distribution.

\begin{table}
\caption{Details of jump parameters used in final global MCMC analysis}
\label{tab:jump}
\begin{tabular}{lrr}
\hline
\hline
\multicolumn{3}{l}{Core parameters} \\
\hline
Parameter & Definition & Prior shape\\
\hline
$T_{0}$ & Transit epoch, HJD - 2450000.0 & Uniform\\ 
$P$ & Orbital period, days & Uniform\\ 
$\Delta F$ & Transit depth, magnitudes & Uniform\\ 
$t_{T}$ & Transit duration, days & Uniform\\ 
$b$ & Impact parameter, R$_{\star}$ & Uniform\\
\teff & Stellar effective temperature, K & Gaussian\\
{[Fe/H]} & Stellar abundance & Gaussian\\ 
$K_{1}$ & RV semi-amplitude, km s$^{-1}$ & Uniform\\ 
\hline
\multicolumn{3}{l}{Optional parameters} \\
\hline
$\sqrt{e} \cos \omega^{a}$ & Eccentricity proxy & Uniform \\
$\sqrt{e} \sin \omega^{a}$ & Eccentricity proxy & Uniform \\
$\sqrt{\vsini } \cos \lambda^{b}$ & Misalignment proxy & Uniform \\
$\sqrt{\vsini } \sin \lambda^{b}$ & Misalignment proxy & Uniform \\
$FWHM_{p}$$^{c}$ & FWHM of planet profile, km s$^{-1}$ & Gaussian \\
$\vsini^{d}$ & Stellar rotation speed, km s$^{-1}$ & Uniform \\
$\lambda^{d}$ & Spin-orbit misalignment angle, $^{\circ}$ & Uniform \\

\hline
\hline
\\
\end{tabular}
\newline {\bf Note:} $^{a}$ parameters used only when a non-circular orbit is allowed. $^{b}$ parameters used when fitting the RM effect. $^{c}$ parameter used when fitting tomographic transit. If $\vsini >> 0$ parameters denoted $^{d}$ are used as jump parameters for tomographic transit fitting.
\end{table}

\begin{table*}
\caption{Table of planetary and stellar parameters found in the global system fit for WASP-92, 93 and 118.}
\label{tab:parameters}
\begin{tabular}{lcrrrl}
\hline
\hline
Parameter & Symbol & WASP-92 & WASP-93 & WASP-118 & Units \\
\hline
General information &&&&& \\
\hline
RA &&                        16h 26m 46.08s           & 00h 37m 50.11s          & 01h 18m 12.12s & J2000 \\
Dec &&                      +51$^\circ$ 02' 28.2'' & +51$^\circ$ 17' 19.5'' & +02$^\circ$ 42' 10.2'' & J2000 \\
V magnitude &&        13.18                            & 10.97                           & 11.02 & \\
K magnitude &&        11.52                            & 9.94                             & 9.79 & \\
\hline 
Spectroscopic analysis results &&&&& \\
\hline
Effective temperature   & \teff              &   6280 $\pm$ 120   & 6700 $\pm$ 100        & 6410 $\pm$ 125             & K \\
Stellar surface gravity            & \logg            &   4.40 $\pm$ 0.12   & 4.5 $\pm$ 0.20          & 4.30 $\pm$ 0.10             &\\
Projected stellar rotation speed             & \vsini            &   5.73 $\pm$ 1.15   & 37 $\pm$ 3                & 9.68 $\pm$ 1.14             & \kms \\
Macroturbulence        & \mactrb        &   4.97 $\pm$ 0.73   & 6.95 $\pm$ 0.73        & 5.77 $\pm$ 0.73             & \kms \\
Microturbulence         & \mictrb         &   1.05 $\pm$ 0.10   & --                                & 1.00 $\pm$ 0.06             & \kms \\
Metallicity                    &{[Fe/H]}        &   0.00 $\pm$ 0.14   & 0.07 $\pm$ 0.17        & 0.16 $\pm$ 0.11             &\\
Lithium abundance    &$\log A$(Li)  &    2.70 $\pm$ 0.09  & $<$ 1.14 $\pm$ 0.09 & $<$   1.21 $\pm$ 0.09   &\\
Stellar mass       &  $M_*$   &   1.20 $\pm$ 0.10   & 1.30 $\pm$ 0.11        & 1.03 $\pm$ 0.08             & $M_{\sun}$ \\
Stellar radius      &   $R_*$                  &   1.14 $\pm$ 0.18   & 1.06 $\pm$ 0.26        & 1.17 $\pm$ 0.16             & $R_{\sun}$ \\
Spectral Type     & &   F7                         & F4                               & F6                                   &\\
Distance      & d &   530 $\pm$ 90   & 250 $\pm$ 60            & 250 $\pm$ 35                 & pc \\ 
\hline 
MCMC jump parameters &&&&& \\
\hline 
Transit epoch (HJD-2450000.0)    & $T_0$                       & $6381.28340 \pm 0.00027$    & $6079.56420 \pm 0.00045$      & $6787.81423 \pm 0.00062$ & days \\           
Orbital period                                & $P$                           & $2.1746742 \pm 0.0000016$  & $2.7325321 \pm 0.0000020$    & $4.0460435 \pm 0.0000044$ & days \\           
Planet/star area ratio                     & $(R_{\rm p}/R_*)^2$   & $0.01254 \pm 0.00029$         & $0.01097 \pm 0.00013$           & $0.00755 \pm 0.00019$ & \\
Transit duration                             & $t_T$                         & $0.1153 \pm 0.0012$              & $0.0931 \pm 0.0010$                & $0.2002 \pm 0.0019$ & days \\           
Impact parameter                         & $b$                            & $0.608 \pm 0.043$                & $0.9036 \pm 0.009$                    & $0.16 \pm 0.10$ & $R_*$ \\           
Stellar effective temperature        & \teff                           & $6258.\pm 120$                       & $6696. \pm 101.$                       & $6420. \pm 121.$ & K \\ 
Metallicity                        & {[Fe/H]}                        & $0.00 \pm 0.14$                      & $0.06 \pm 0.17$                        & $0.16 \pm 0.11$ & \\ 
Stellar reflex velocity                   & $K_1$                         & $0.1116 \pm 0.0092$  &            $0.174 \pm 0.034$                      & $0.0546 \pm 0.0019$ & km s$^{-1}$ \\ 
\hline 
Deduced stellar parameters &&&&& \\
\hline
Stellar density                               & $\rho_*$                     & $0.495 \pm 0.055$                  & $0.376 \pm 0.028$                     & $0.271 \pm 0.012$ & $\rho_\odot$ \\           
Stellar surface gravity                  & $\log g_*$                  & $4.259 \pm 0.031$                   & $4.197 \pm 0.019$                    & $4.100 \pm 0.012$ & (cgs) \\           
Projected stellar rotation speed   & $v\sin i$                      & 5.73 (fixed)                              & 37.0 (fixed)                               & 9.68 (fixed) & km s$^{-1}$ \\           
Stellar radius                                & $R_*$                        & $1.341 \pm 0.058$                   & $1.524 \pm 0.040$                      & $1.696 \pm 0.029$ & $R_\odot$ \\           
Stellar mass                                 & $M_*$                         & $1.190 \pm 0.037$                  & $1.334 \pm 0.033$                     & $1.320 \pm 0.035$ & $M_\odot$ \\
Scaled stellar radius                     & $R_*/a$                     & $0.1790 \pm 0.0067$              & $0.1684 \pm 0.0037$                  & $0.1445 \pm 0.0022$ & \\
Centre-of-mass velocity               & $\gamma$                 & $-29.1683 \pm 0.0067$          & $-2.785 \pm 0.029$                     & $5.8826 \pm 0.0038$ & km s$^{-1}$ \\
\hline 
Deduced planet parameters &&&&& \\
\hline
Orbital separation                        & $a$                             & $0.03480 \pm 0.00036$        & $0.04211 \pm 0.00035$            & $0.05453 \pm 0.00048$ & AU \\           
Orbital inclination                        & $i$                               & $83.75 \pm 0.69$                   & $81.18 \pm 0.29$                       & $88.70 \pm 0.90$ & $^\circ$ \\           
Orbital eccentricity                       & $e$                              & 0 (fixed)                                  & 0 (fixed)                                     & 0 (fixed) & \\           
Argument of periastron                & $\omega$                    &  0 (fixed)                                & 0 (fixed)                                       & 0 (fixed) & $^\circ$ \\           
Spin-orbit misalignment angle     & $\lambda$                  & 0 (fixed)                                  & 0 (fixed)                                     & 0 (fixed) & $^\circ$ \\ 
Planet radius                               & $R_{\rm p}$                & $1.461 \pm 0.077$                  & $1.597 \pm 0.077$                      & $1.440 \pm 0.036$ & $R_{\rm J}$ \\           
Planet mass                                & $M_{\rm p}$                & $0.805 \pm 0.068$                  & $1.47 \pm 0.29$                          & $0.514 \pm 0.020$ & $M_{\rm J}$ \\           
Planet surface gravity                 & $\log g_{\rm p}$          & $2.939 \pm 0.056$                 & $3.120 \pm 0.093$                     & $2.757 \pm 0.040$ & (cgs) \\           
Planet density                             & $\rho_{\rm p}$             & $0.260 \pm 0.044$              & $0.360 \pm 0.084$                       & $0.175 \pm 0.018$ & $\rho_{\rm J}$ \\           
Planetary equilibrium temperature & $T_{\rm P}$           & $1871. \pm 56.$                  & $1942. \pm 38.$                        & $1729. \pm 36.$ & K \\
Roche limit                                  & $a_{R}$                      & $0.0171 \pm 0.0010$        & $0.0157 \pm 0.0015$                 & $0.0204 \pm 0.0009$ & AU \\
Roche separation                       & $a / a_{R}$                  & $2.038 \pm 0.026$           &  $2.685 \pm 0.245$                    &  $2.671 \pm 0.099$ &  \\           
\hline
\hline
\\
\end{tabular}
\newline {\bf Note on spectroscopic parameters:} Mass and Radius estimate using the \cite{torres2010} calibration. Spectral Type estimated from \teff using the table in \cite{gray2008}. Abundances are relative to the solar values obtained by \cite{asplund2009}.
{\bf Note on MCMC parameters:} The Roche limit is $a_{R} = 2.16 R_{\rm p} (M_* / M_{\rm p})^{1/3}$, as defined in \citet{ford2006}.
\end{table*}

This MCMC parameter fitting analysis was performed for each of WASP-92, 93 and 118 using the WASP photometry, follow up photometric lightcurves and spectroscopic follow up data relevant to each target. The values and uncertainties for stellar properties found in the spectral analysis in Section \ref{sec:stellar} were used as initial values, and the orbital parameters were initialised from a box least squares fit of the WASP photometry.

The final parameters found in the analysis are quoted in Table \ref{tab:parameters}. For each system, a circular orbit was assumed for a preliminary fit, which was then allowed to be eccentric in a subsequent fit. In each case, there was not enough statistical justification that adding the eccentricity parameters improved the fit of the system, which was determined using the Bayesian Information Criterion (BIC) \citep{schwarz1978,kass1995}. Thus the circular orbital solutions were used as the final results.

\section{Results and discussion}
\label{sec:discu}


\begin{figure}
\includegraphics[width=0.48\textwidth]{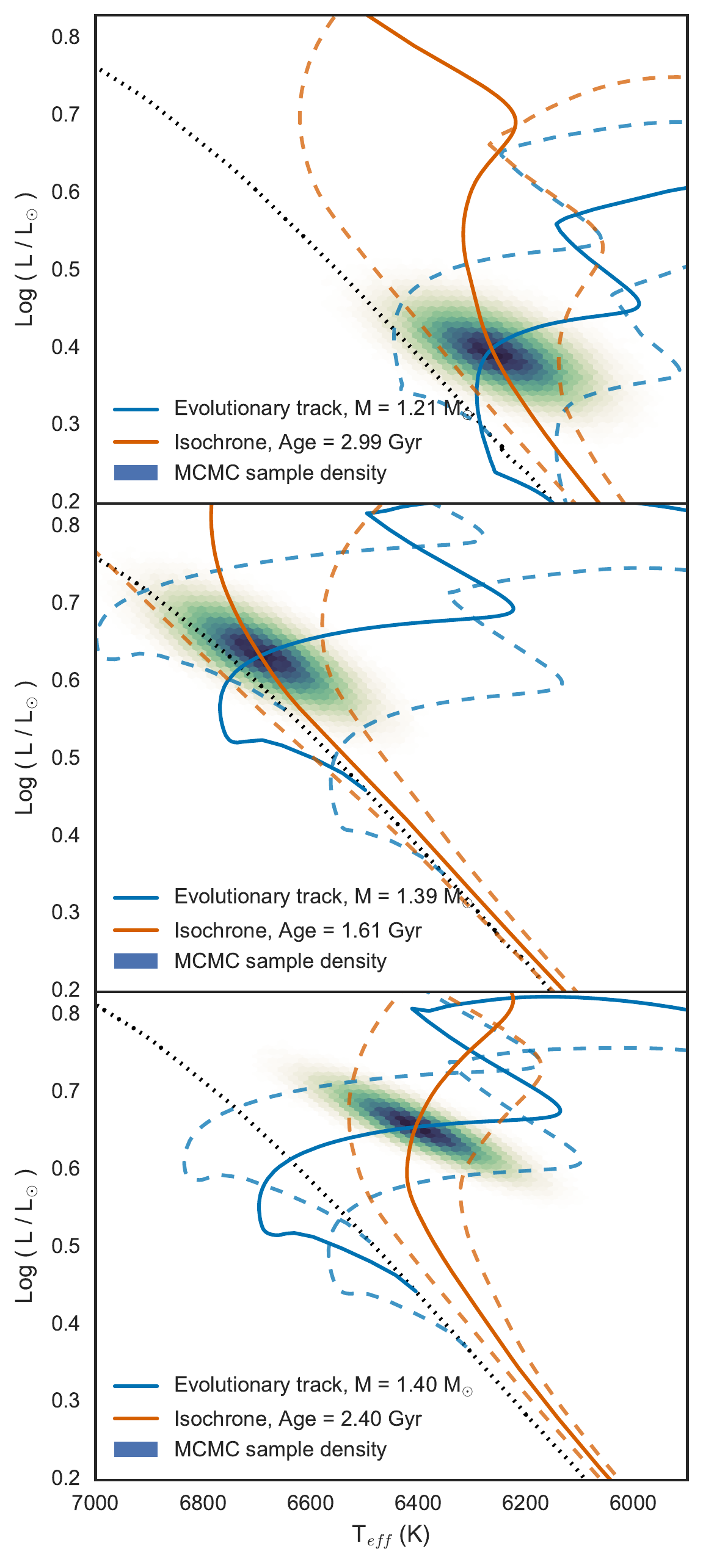}
\caption{Results of the \textsc{Bagemass} MCMC analysis for WASP-92 (upper plot); WASP-93 (middle), and WASP-118 (lower). For each of the plots, the dotted black line is the ZAMS. The solid blue line is the evolutionary track for the mass found, and the dashed tracks either side are for the 1-$\sigma$ error of the mass. The solid orange line is the isochrone for the stellar age found, with the 1-$\sigma$ error denoted by dashed lines in the same colour. The density of MCMC samples is shown in the colour scale of the posterior distribution plotted.}
\label{fig:evo}
\end{figure}

\subsection{WASP-92 system}

WASP-92b is a {\bf$\sim0.81$} $M_{\rm J}$ planet in a 2.17 days orbit around an F7 spectral type star. The \textsc{Bagemass} tool presented in \citet{maxted2015} was used with the observed measurements for {[Fe/H]}, \teff and $\rho_*$ to estimate the age and mass of WASP-92, which were found to be $2.99\pm1.03$ Gyr and $1.21\pm0.06$ $M_\odot$. The posterior distributions of this analysis can be seen in the upper plot of Figure \ref{fig:evo} with the associated stellar evolution tracks and isochrones for the optimal mass and age found by the \textsc{Bagemass} tool. The tracks used by the tool were calculated from the \textsc{GARSTEC} code \citep{weiss2008}. These results are in agreement with the gyrochronological age presented in Section \ref{sec:stellar92}, but much older than the age suggested by the lithium abundance observed in the spectra.

When the solution for the system was allowed to include eccentricity, the value for $e$ found was $0.084^{+0.118}_{-0.060}$, which affected each of the reported parameters by $<$5\%. The BIC for the eccentric solution was 19.1, and 5.7 for the circular solution. Given the low value for eccentricity found and a $\Delta$BIC\footnotemark \footnotetext{where $\Delta$BIC = BIC$_{ecc}$ - BIC$_{circ}$} $>$10, which \citet{kass1995} indicates is very strong evidence against additional free parameters, the orbit of WASP-92b can be assumed to be circular. The parameters found in the fit for a circular orbit were used as the final parameters.

\subsubsection{Tidal evolution}

Using the parameters of the orbital solution, the tidal stability of the system can be investigated. The majority of hot Jupiters observed are in Darwin-unstable orbits \citep{darwin1879}, where the planet is migrating towards the Roche limit of the system and tidal disruption of the planet \citep{matsumura2010}. No stable orbits exist if the total angular momentum of the system, $L_{tot}$ is below a critical angular momentum, $L_{c}$,
\begin{equation}
L_{c} = 4 \left( \frac{G^{2}}{27} \frac{M_{*}^{3}M_{p}^{3}}{M_{*}+M_{p}} \left( C_{*} + C_{p}\right) \right)^{\frac{1}{4}} \,\, ,
\end{equation} 
where $C_{*}$ and $C_{p}$ are the moments of inertia of the star and planet respectively \citep{counselman1973,hut1980}. The total angular momentum is defined as 
\begin{equation}
L_{tot} = L_{orb} + C_{*}\omega_{*} + C_{p}\omega_{p} \, \, ,
\end{equation}
where $L_{orb} = M_{*}M_{p}\sqrt{ \frac{Ga(1-e^{2})}{M_{*} + M_{p}} }$. For WASP-92, $L_{tot} / L_{c} \sim 0.67$, which indicates that the planetary orbit is unstable and is continuing to migrate inwards towards the Roche limit, which is the case for most hot Jupiters. 

The rate of this migration can be estimated by,
\begin{equation}
t_{\textrm{remain}} = \frac{2 Q'_{*,0}}{117n} \frac{M_{*}}{M_{p}} \left( \frac{a}{R_{*}} \right)^{5} \,\, ,
\end{equation}
as presented in \citet{brown2011} for slowly rotating stars, where $Q'_{*,0}$ is the current tidal quality factor for the star and $n$ is the orbital frequency. If $Q'_{*,0}$ is set to $10^8$\footnotemark \footnotetext{The use of $Q'_{*,0} = 10^{8}$ is suggested in \citet{penev2011}.}, the spiral in time ($t_{\textrm{remain}}$) is approximately 16 Gyr, which is significantly longer than the remaining lifetime of the host star. For lower values of $Q'_{*}$, the decay timescale remains too large for changes in the orbital period to be observable - a period change of $>60$s will occur after $\sim 20$ Myr for $Q'_{*,0} = 10^{8}$ when solely accounting for the effects of tidal orbital decay.

\subsection{WASP-93 system}

WASP-93b is a {\bf$\sim1.47$} $M_{\rm J}$ planet in a 2.73 days orbit around an F4 spectral type star. \textsc{Bagemass} was used for the WASP-93 system, resulting in age and mass estimates of $1.61\pm0.48$ Gyr and $1.39\pm0.08$ $M_\odot$. This age estimate is just beyond the upper 1-$\sigma$ uncertainty for the age derived from gyrochronology, and below the estimate of several Gyr from the lithium abundance.

When the eccentricity parameters were included in the orbital solution, the value for $e$ found was $0.012^{+0.035}_{-0.008}$, and eccentric BIC was 168.3 and circular BIC was 147.4. The orbit is assumed to be circular, as the value for $e$ is small, and the $\Delta$BIC$>$10.

\begin{figure}
\includegraphics[width=0.49\textwidth]{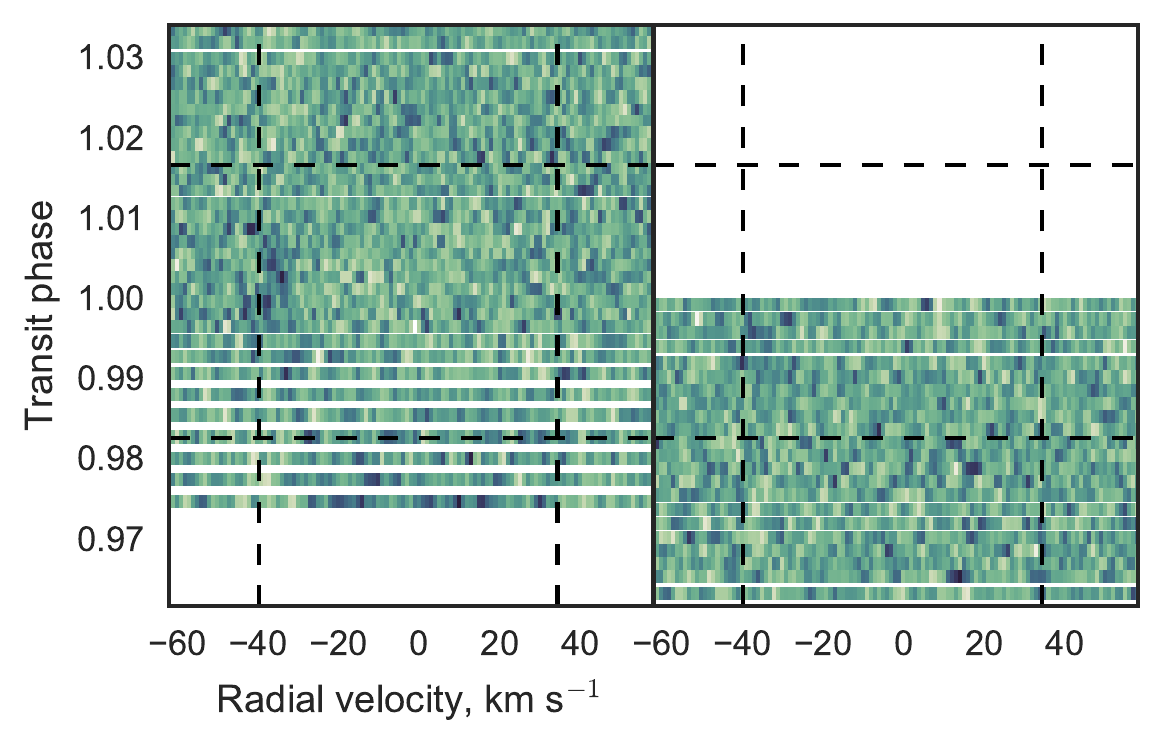}
\includegraphics[width=0.49\textwidth]{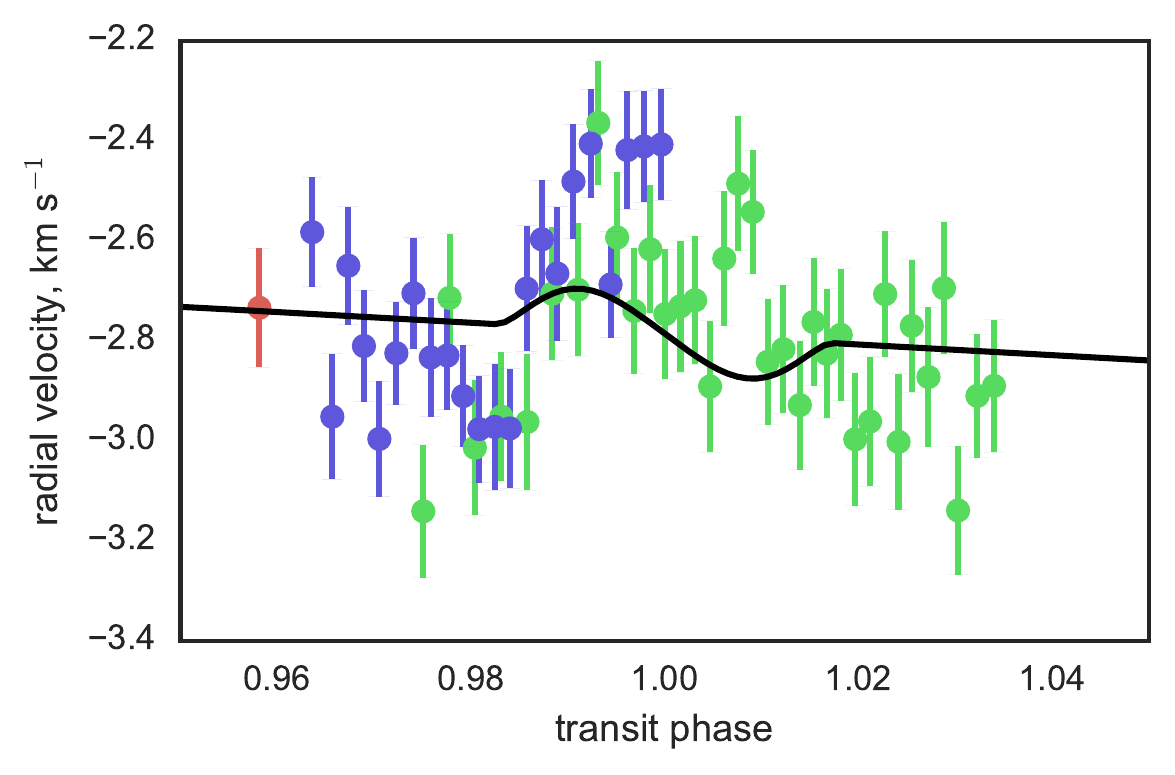}
\caption{Upper plot showing the time series of CCF residuals after the subtraction of the average CCF profile with reference to transit phase for each of the two attempts to observe spectra during transit. The vertical dashed lines show the width of the CCF profile centred on the centre-of-mass velocity of the system. The horizontal dashed lines show beginning of transit ingress and egress. The trail of the transit of WASP-93b is visible near the blue-shifted limb of the CCF.
Lower plot showing the RV measurements around the transit phase. The colours used correspond to those in the lower plot of Figure \ref{fig:wasp93}, and the green and blue RV points match the CCFs shown in the left and right of the upper plot respectively. The RV and RM effect model overplotted in black is for the case that the planetary orbit and the stellar spin are aligned.}
\label{fig:w93tomography}
\end{figure}

In Section \ref{sec:spec} it was mentioned that two attempts were made to collect a time series of spectra during transits of WASP-93b. Due to uncertainty in the transit ephemeris at the time of these observations, the measurements are not well centred on the mid-transit time. The CCFs for these observations were reduced as described in \citet{cameron2010}, and included in the fit, which now includes the orbital obliquity and FWHM of the planet signal in the CCF as jump parameters. The MCMC code did not converge on a solution when including the CCFs to determine orbital obliquity, which indicates that the SOPHIE spectra were not of a high enough precision to determine the misalignment of the transit of WASP-93b. A value for the orbital obliquity was also not found when the $v \sin i$ was fixed to the value found in the global fit presented in Table \ref{tab:parameters}, $37.0$ km s$^{-1}$. 

The maximum amplitude of the RV anomaly produced by the RM effect is defined as,
\begin{equation}
\Delta RV_{RM} = \sqrt{1 - b^{2}}\left( \frac{R_{p}}{R_{\star}} \right) ^{2} v \sin i \,\, ,
\end{equation}
which equates to $\sim 0.18$ km s$^{-1}$ for WASP-93. The average 1-$\sigma$ uncertainty for the first series of spectra is 0.13 km s$^{-1}$ and 0.11 km s$^{-1}$ for the second, which will make the effect difficult to detect above the noise in the CCFs. The lower plot in Figure \ref{fig:w93tomography} shows the RV measurements calculated for the CCFs shown in the upper plot with reference to transit phase. The plotted RM effect model is for an aligned orbit, for the data shown, it is unclear what the best fit shape of the RM effect curve should be, which further indicates that higher quality data is required to fully determine the orbital obliquity of the WASP-93 system. 

The upper plot in Figure \ref{fig:w93tomography} shows the time series of the residuals of the CCFs after the subtraction of the average CCF shape, in which a signature of planetary transit is visible near the blue shifted limb of the CCFs, particularly in the first set of data. The effect is predominantly visible during the centre of the transit, where the most starlight is occulted by the planet. Given that none of the red shifted limb of the CCF is occulted by the observed planet signal, it is expected that WASP-93b has an orbit almost entirely mutually misaligned with the stellar spin axis.

\subsubsection{Blend scenario}

Given the shape of the transit appears to be almost v-shaped, it is important to investigate whether a background blended eclipsing binary or hierarchical triple system could be producing the eclipse signal observed in the lightcurve of WASP-93 and the RV variation detected. The AO image shows a blended stellar companion separated by $0.69\pm0.01$ arcsec, which is fainter by more than 3 magnitudes, so would not be able to produce the transit depth observed, even if the star was fully occulted by a non-emitting body. The AO imaging also shows no other stellar bodies within 0.3 arcsec of the stellar core of WASP-93 which would be bright enough to produce an eclipse mimicking a planet transiting WASP-93.

Whilst the transit signal could be created by a chance-aligned background eclipsing binary which is not resolved in the AO images, this scenario is extremely unlikely given the small region of space where a bright enough stellar binary system would have to exist in to not be resolved in the AO image. 

The most likely blend scenario would be that WASP-93 is a hierarchical triple system, but this would produce a blended line profile in the observed CCFs. Even with CCFs calculated to $\pm 100$ km s$^{-1}$ from the systemic RV, there is no evidence of an additional line profile. The modulation of the CCF shape during the transit also confirms the planetary nature of the signal, rather than that of a hierarchical triple system.

\subsubsection{Tidal evolution}

\begin{figure}
\includegraphics[width=0.49\textwidth]{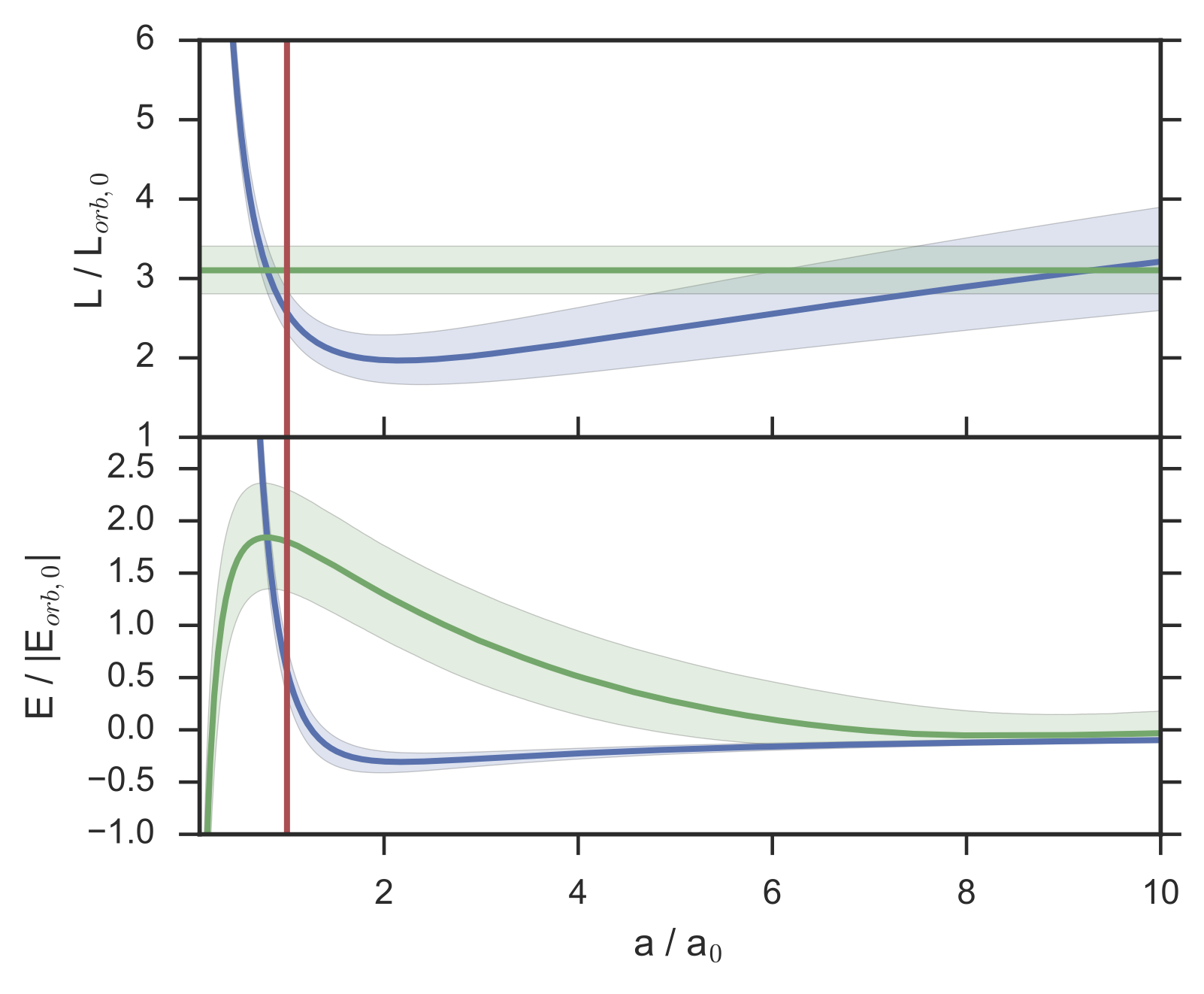}
\caption{Plot showing the tidal equilibrium curves for WASP-93. In the upper plot, the blue line shows the total angular momentum of the system when dual-synchronised for the range of semi-major axis; the green line shows the current total angular momentum with an assumption of spin-orbit alignment, and the red line shows the current separation of the star and planet. Each of the angular momenta are scaled with the current total orbital angular momentum, L$_{orb,0} = 2.986\times 10^{42}$ kg m$^{2}$/s. The lower plot shows the curves for the total energy in the system, where the blue line shows the total orbital and rotational energy for the system when dual synchronised for the range of orbital separation; the green line shows the total energy when angular momentum is conserved, and the red line shows the current separation. Each of the energies are scaled with the current orbital energy, E$_{orb,0} = -3.976\times 10^{37}$ kg m$^{2}$/s$^{2}$. The 1-$\sigma$ uncertainties plotted are calculated from the output chains of the global MCMC analysis of the system.}
\label{fig:w93tidal}
\end{figure}

Unlike most hot Jupiters, WASP-93 has a value of $L_{tot} > L_{c}$, which indicates that an orbit exists for this system where tidal equilibrium can be reached -- in this case $L_{tot} / L_{c} = 1.57 \pm{0.16}$. Tidal equilibrium is characterised by the minimisation of total orbital and spin energy, constrained by the conservation of angular momentum. Figure \ref{fig:w93tidal} shows where the current angular momentum of the system (in green) intersects with the angular momentum curve for the dual synchronous state (in blue), which is where stable orbits exist. The current orbital separation of the star and planet (in red) is between the inner and outer equilibrium states. The inner equilibrium is unstable, which can be seen in the lower plot demonstrating the total energy of the system, where that equilibrium denotes a maximum in energy. The energy in the plot has been defined as
\begin{equation}
E_{tot} = -\frac{G M_{*} M_{p} }{2a} + \frac{1}{2} C_{*} \omega_{*}^{2} + \frac{1}{2} C_{p} \omega_{p}^{2} \, \, ,
\end{equation}
where the first term is the orbital energy, and the remaining terms are the rotational energy of the star and planet respectively. In the case of pseudo-synchronisation, the orbital frequency, $n = \omega_{p}$, and the value for $\omega_{*}$ is determined by angular momentum conservation (green line) or synchronisation with the orbital frequency (blue line).

If the planetary orbit and the stellar spin are aligned and since angular momentum is conserved, the WASP-93 system will migrate along the path of constant $L$ until it reaches the outer equilibrium, which is at an energy minimum and is a Darwin stable dual-synchronised orbit. The timescale for this outwards migration can be approximated using a numerical integration of Equation 18 from \citet{matsumura2010}\footnotemark \footnotetext{$\frac{da}{dt} = \frac{9}{Q'_{*,0}}\frac{M_{p}}{M_{*}}\frac{R^{5}_{*}}{a^{4}} \left( \frac{a_{0}}{a} \right)^{3/2} (\omega_{*,0}\cos{\epsilon_{*,0}} - n)$, where $\epsilon_{*,0}$ is the current spin orbit misalignment.}. Using this integration, the timescale until the equilibrium orbit is reached is $\sim 2 \times 10^{7}$ Gyr for $Q'_{*,0}=10^{8}$, which is far longer than the remaining lifetime of the star even for lower values of $Q'_{*,0}$. The timescale for an observable change in the orbital period ($>60$s) is also too long to detect at $\sim 50$ Myr for $Q'_{*,0}=10^{8}$.

Since the tomographic analysis indicates that the orbit of the planet is significantly misaligned with the spin axis of the host star, the current total angular momentum of the system in the plane of the stellar rotation is lower than if the system were aligned. If the orbits are misaligned, beyond $\lambda \sim 30^{\circ}$, which is indicated by the tomographic signal, then either $L_{tot} < L_{c}$ or the current separation of the star and the planet would mean that the planet is tidally migrating towards its host star.

\subsection{WASP-118 system}

\begin{figure}
\includegraphics[width=0.45\textwidth]{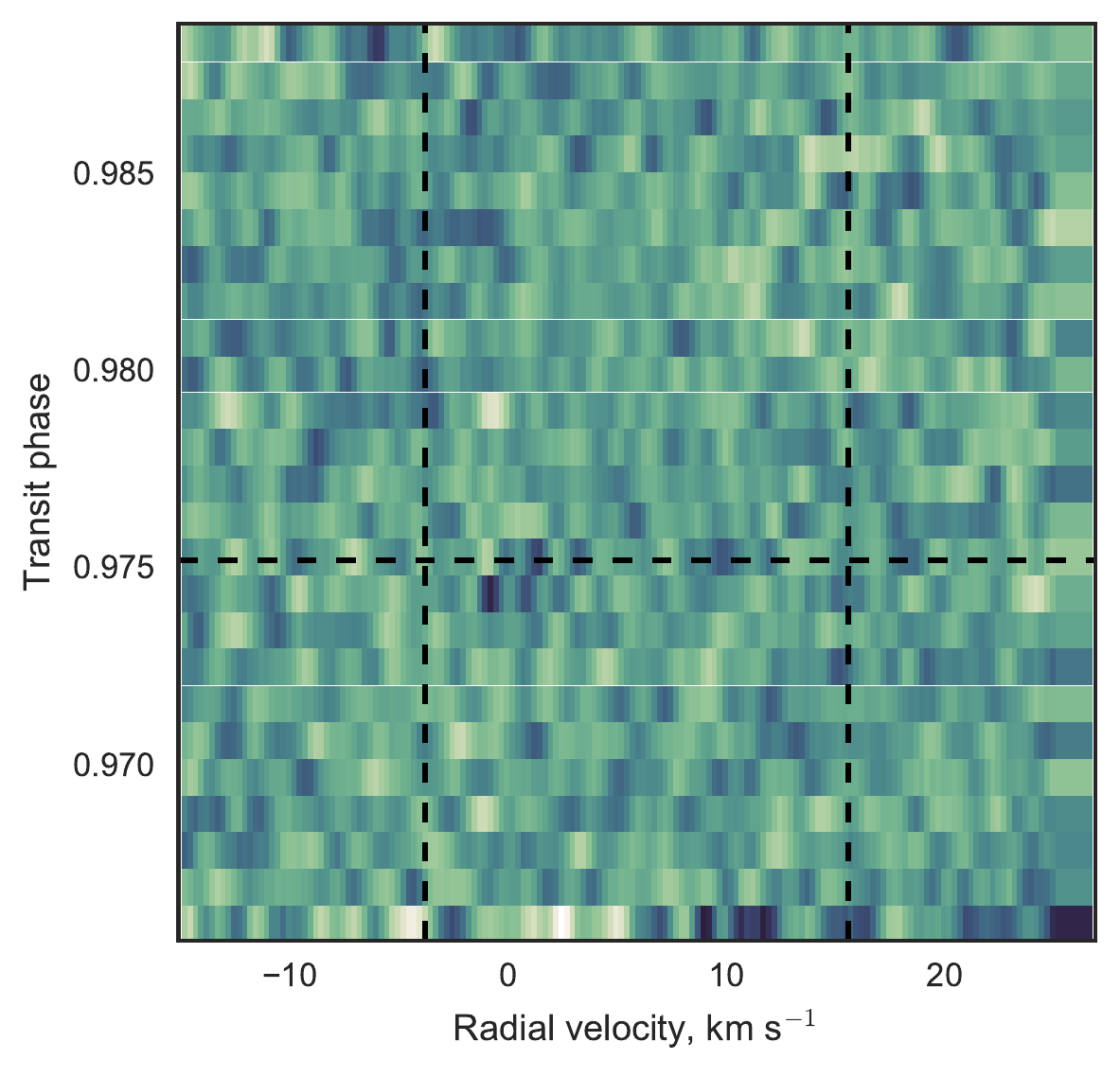}
\includegraphics[width=0.45\textwidth]{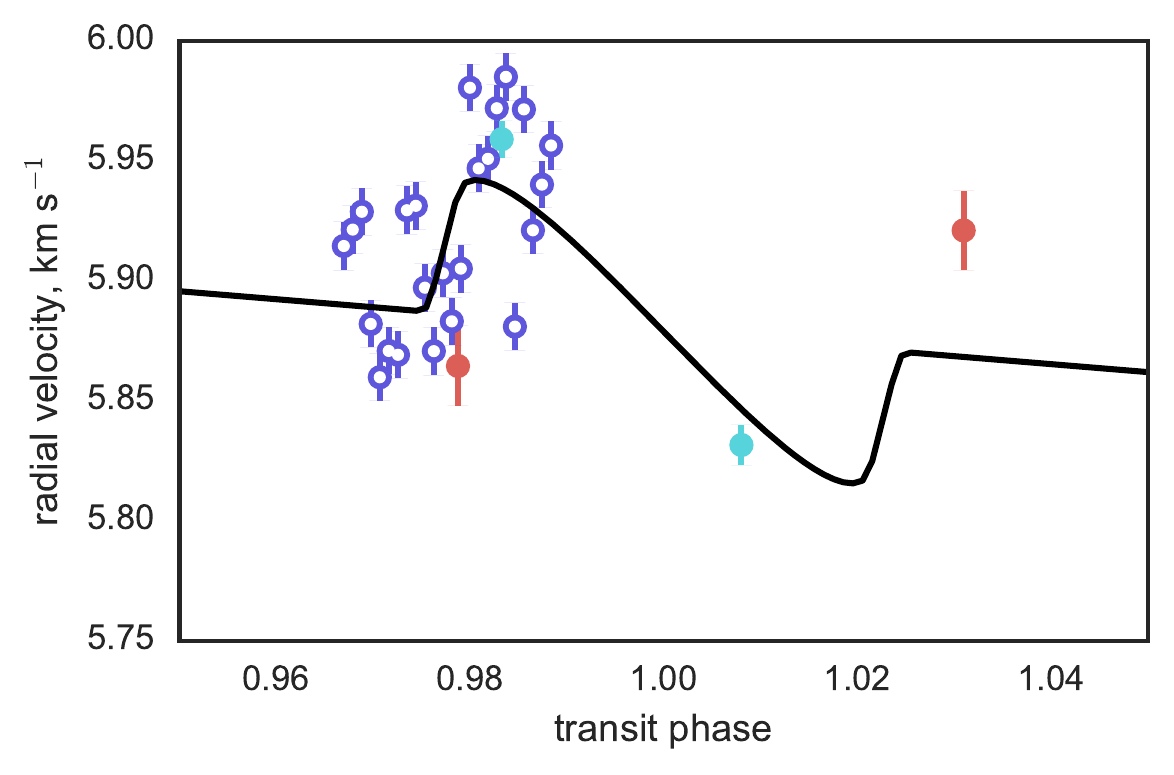}
\caption{Upper plot showing the time series of CCF residuals after the subtraction of the average CCF shape with reference to transit phase the attempt to observe spectra throughout a transit, where the trail of the transit of WASP-118b is not visible. Lower plot showing the RV measurements around the transit phase. The colours used correspond to those in the lower plot of Figure \ref{fig:wasp118}, and the purple RV points match the CCFs shown in the upper plot.}
\label{fig:w118detail}
\end{figure}

WASP-118b is a {\bf$\sim 0.52$} $M_{\rm J}$ planet in a 4.04 days orbit around an F6 spectral type star. \textsc{Bagemass} was also used to obtain mass and age estimates for WASP-118, which gave $2.38\pm 0.38$ Gyr and $1.40\pm 0.05$ $M_\odot$. The results of the \textsc{Bagemass} modelling are shown in the lower plot in Figure \ref{fig:evo}, which shows that WASP-118 is slightly evolved. This result is in agreement with the spectral analysis for this system in Section \ref{sec:stellar118}.

Since the weather on La Palma was poor for the spectral observations on the night beginning 1st October 2015, the uncertainties calculated by the DRS appear lower than would be expected for these conditions. The time series of CCFs collected can be seen in the upper plot of Figure \ref{fig:w118detail}, which highlights the varying observing conditions, and the effect of the 79\% illuminated moon an average of 42$^{\circ}$ away from the star during the observations. The CCFs with the average CCF profile subtracted showed no sign of the planet occulting part of the line profile during transit, which is not surprising given the noisy data.

The scatter in the RV measurements calculated can also be clearly seen in the lower plot in Figure \ref{fig:w118detail}, which shows the measured RVs with respect to transit phase. Due to the concern about the reliability of these measurements, which could bias the results, the global fit was completed and compared with the inclusion and exclusion of this data set. 

With the transit series of HARPS-N RVs excluded when the solution was allowed to be non-circular, the value for $e$ was $0.089^{+0.057}_{-0.047}$. The associated BICs were 20.3 for a circular solution, and 67.0 for an eccentric solution. If the extra RV points are included in the global fit, $e=0.132^{+0.105}_{-0.080}$, and the BIC for a circular solution is 50.3, whereas for a non-circular solution the BIC is 520.4. In each case, the $\Delta$BIC $=$ BIC$_{ecc}$ - BIC$_{circ}$ $>$ 10, which is strong evidence against including the two extra free parameters for eccentricity, so the system is assumed to be circular.

The scatter of the transit series of HARPS-N RVs around the fitted model, which is not accounted for with the uncertainties quoted, would not be expected to be caused by stellar activity on the timescale of a few of hours. The data points have been excluded from the fit used for the final parameters quoted, to prevent the underestimated uncertainties from biasing the parameters presented in this work. 

The distribution of the RV measurements taken during the transit suggest that the orbit of WASP-118b is aligned with respect to the stellar spin axis, which suggests a disc-migration origin for the planet and makes the system a strong target for looking for companion planets, such as those found for WASP-47 \citep{becker2015,neveuvanmalle2015}. Were any companion planets around WASP-118 assumed to have orbits co-planar with the orbit of WASP-118b, planets up to 0.45 AU\footnotemark \footnotetext{Using the relation $a_{t} = {R_*}/{\sin(|90-i|)}$, where $a_{t}$ is the furthest orbital separation where transits will occur for the given inclination of the orbit.} away from the host star would also transit, which will be detectable with K2 depending on the radii of the planets. 

\subsubsection{Tidal evolution}

\begin{figure}
\includegraphics[width=0.49\textwidth]{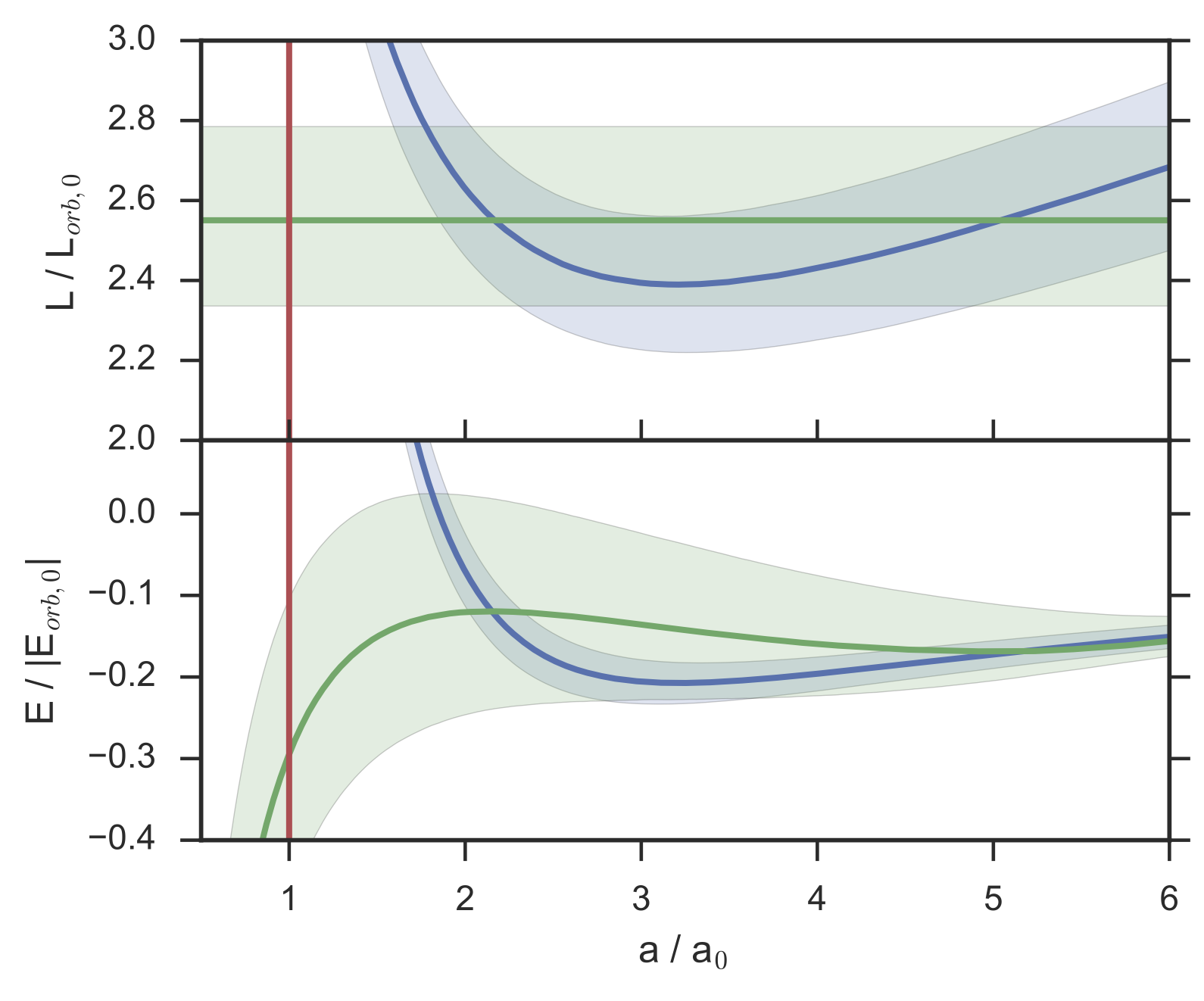}
\caption{Plot showing the tidal equilibrium curves for WASP-118, which are presented in the same way as Figure \ref{fig:w93tidal}. The upper plot is scaled with orbital angular momentum, L$_{orb,0} = 1.164\times 10^{42}$ kg m$^{2}$/s, and the lower plot is scaled with orbital energy, E$_{orb,0} = -1.046\times 10^{37}$ kg m$^{2}$/s$^{2}$.}
\label{fig:w118tidal}
\end{figure}

WASP-118 also has $L_{tot} > L_{c}$ ($=1.07\pm 0.08$), thus the system very likely has enough angular momentum to exist in a tidal equilibrium with the same assumptions about the system geometry used above. In order to test whether the system is evolving to a stable orbit, the tidal equilibrium curves for WASP-118 are plotted in Figure \ref{fig:w118tidal} as for WASP-93. The red line denoting the current orbital separation is clearly inside the inner equilibrium state, which indicates the system cannot reach a stable orbit. WASP-118b will follow the path of constant $L$ towards the local lowest energy state, which is tidal migration inwards towards the Roche limit of the star.

Given the relatively high $\vsini$ of WASP-118, the expression for $t_{\textrm{remain}}$ used for WASP-92b is not an appropriate approximation in this case. In order to estimate the spiral in time for the system, the equation for $\frac{da}{dt}$ used for WASP-93b from \citet{matsumura2010} was numerically integrated from the current orbital separation to the stellar radius. In order to construct this integral, it is approximated that $e$=0, and that the planetary spin is synchronised with the orbital period, which are valid assumptions for a short period giant planet system. The spin and orbit are also assumed to be aligned, which is implied from the in transit RV measurements for WASP-118. The result of the calculation is $t_{\textrm{remain}} \sim 150$ Gyr when $Q'_{*,0}=10^{8}$. As for WASP-92b, the spiral in timescale of the planet is too long to be observable even with a higher tidal efficiency (lower $Q'_{*,0}$.)

\section*{Acknowledgments}

KH and ACC acknowledge support from STFC doctoral training grant ST/M503812/1 and grant ST/M001296/1 respectively. We extend our thanks to the staff of the ING for their continued support of SuperWASP. WASP-South is hosted by the South African Astronomical Observatory, and we are grateful for their ongoing support and assistance. The Euler Swiss telescope, installed at La Silla, is supported by the Swiss National Science Foundation. We are all grateful to ESO and its La Silla staff for their continuous support. TRAPPIST is a project funded by the Belgian Fund for Scientific Research (Fonds National de la Recherche Scientifique, F. R. S.-FNRS) under grant FRFC 2.5.594.09.F, with the participation of the Swiss National Science Foundation. L. Delrez acknowledges the support of the F. R. I. A. fund of the FNRS. M. Gillon and E. Jehin are FNRS Research Associates. The research leading to these results has received funding from the European Community's Seventh Framework Programme (FP7/2013-2016) under grant agreement number 312430 (OPTICON). The William Herschel Telescope is operated on the island of La Palma by the Isaac Newton Group in the Spanish Observatorio del Roque de los Muchachos of the Instituto de Astrofísica de Canarias.

\bibliographystyle{mnras}
\bibliography{wasp_reference}

\begin{thebibliography}{}
\makeatletter
\relax
\def\mn@urlcharsother{\let\do\@makeother \do\$\do\&\do\#\do\^\do\_\do\%\do\~}
\def\mn@doi{\begingroup\mn@urlcharsother \@ifnextchar [ {\mn@doi@}
  {\mn@doi@[]}}
\def\mn@doi@[#1]#2{\def\@tempa{#1}\ifx\@tempa\@empty \href
  {http://dx.doi.org/#2} {doi:#2}\else \href {http://dx.doi.org/#2} {#1}\fi
  \endgroup}
\def\mn@eprint#1#2{\mn@eprint@#1:#2::\@nil}
\def\mn@eprint@arXiv#1{\href {http://arxiv.org/abs/#1} {{\tt arXiv:#1}}}
\def\mn@eprint@dblp#1{\href {http://dblp.uni-trier.de/rec/bibtex/#1.xml}
  {dblp:#1}}
\def\mn@eprint@#1:#2:#3:#4\@nil{\def\@tempa {#1}\def\@tempb {#2}\def\@tempc
  {#3}\ifx \@tempc \@empty \let \@tempc \@tempb \let \@tempb \@tempa \fi \ifx
  \@tempb \@empty \def\@tempb {arXiv}\fi \@ifundefined
  {mn@eprint@\@tempb}{\@tempb:\@tempc}{\expandafter \expandafter \csname
  mn@eprint@\@tempb\endcsname \expandafter{\@tempc}}}

\bibitem[\protect\citeauthoryear{{Albrecht} et~al.,}{{Albrecht}
  et~al.}{2012}]{albrecht2012}
{Albrecht} S.,  et~al., 2012, \mn@doi [\apj] {10.1088/0004-637X/757/1/18},
  \href {http://adsabs.harvard.edu/abs/2012ApJ...757...18A} {757, 18}

\bibitem[\protect\citeauthoryear{{Anderson} et~al.,}{{Anderson}
  et~al.}{2011}]{anderson2011}
{Anderson} D.~R.,  et~al., 2011, \mn@doi [\apjl] {10.1088/2041-8205/726/2/L19},
  \href {http://adsabs.harvard.edu/abs/2011ApJ...726L..19A} {726, L19}

\bibitem[\protect\citeauthoryear{{Asplund}, {Grevesse}, {Sauval}  \&
  {Scott}}{{Asplund} et~al.}{2009}]{asplund2009}
{Asplund} M.,  {Grevesse} N.,  {Sauval} A.~J.,   {Scott} P.,  2009, \mn@doi
  [\araa] {10.1146/annurev.astro.46.060407.145222}, 47, 481

\bibitem[\protect\citeauthoryear{{Baranne} et~al.,}{{Baranne}
  et~al.}{1996}]{baranne1996}
{Baranne} A.,  et~al., 1996, \aaps, \href
  {http://adsabs.harvard.edu/abs/1996A%26AS..119..373B} {119, 373}

\bibitem[\protect\citeauthoryear{{Barnes}}{{Barnes}}{2007}]{barnes2007}
{Barnes} S.~A.,  2007, \mn@doi [\apj] {10.1086/519295}, 669, 1167

\bibitem[\protect\citeauthoryear{{Becker}, {Vanderburg}, {Adams}, {Rappaport}
  \& {Schwengeler}}{{Becker} et~al.}{2015}]{becker2015}
{Becker} J.~C.,  {Vanderburg} A.,  {Adams} F.~C.,  {Rappaport} S.~A.,
  {Schwengeler} H.~M.,  2015, \mn@doi [\apjl] {10.1088/2041-8205/812/2/L18},
  \href {http://adsabs.harvard.edu/abs/2015ApJ...812L..18B} {812, L18}

\bibitem[\protect\citeauthoryear{{Bouchy} et~al.,}{{Bouchy}
  et~al.}{2009}]{bouchy2009}
{Bouchy} F.,  et~al., 2009, \mn@doi [\aap] {10.1051/0004-6361/200912427}, \href
  {http://adsabs.harvard.edu/abs/2009A%26A...505..853B} {505, 853}

\bibitem[\protect\citeauthoryear{{Brown}, {Collier Cameron}, {Hall}, {Hebb}  \&
  {Smalley}}{{Brown} et~al.}{2011}]{brown2011}
{Brown} D.~J.~A.,  {Collier Cameron} A.,  {Hall} C.,  {Hebb} L.,   {Smalley}
  B.,  2011, \mn@doi [\mnras] {10.1111/j.1365-2966.2011.18729.x}, \href
  {http://adsabs.harvard.edu/abs/2011MNRAS.415..605B} {415, 605}

\bibitem[\protect\citeauthoryear{Cameron et~al.,}{Cameron
  et~al.}{2006}]{cameron2006}
Cameron A.~C.,  et~al., 2006, \mn@doi [\mnras]
  {10.1111/j.1365-2966.2006.11074.x}, 373, 799

\bibitem[\protect\citeauthoryear{Cameron et~al.,}{Cameron
  et~al.}{2007}]{cameron2007}
Cameron A.~C.,  et~al., 2007, \mn@doi [\mnras]
  {10.1111/j.1365-2966.2007.12195.x}, 380, 1230

\bibitem[\protect\citeauthoryear{{Claret}}{{Claret}}{2000}]{claret2000}
{Claret} A.,  2000, \aap, \href
  {http://adsabs.harvard.edu/abs/2000A%26A...363.1081C} {363, 1081}

\bibitem[\protect\citeauthoryear{{Collier Cameron}, {Bruce}, {Miller}, {Triaud}
   \& {Queloz}}{{Collier Cameron} et~al.}{2010}]{cameron2010}
{Collier Cameron} A.,  {Bruce} V.~A.,  {Miller} G.~R.~M.,  {Triaud}
  A.~H.~M.~J.,   {Queloz} D.,  2010, \mn@doi [\mnras]
  {10.1111/j.1365-2966.2009.16131.x}, \href
  {http://adsabs.harvard.edu/abs/2010MNRAS.403..151C} {403, 151}

\bibitem[\protect\citeauthoryear{{Cosentino} et~al.,}{{Cosentino}
  et~al.}{2012}]{cosentino2012}
{Cosentino} R.,  et~al., 2012, in Ground-based and Airborne Instrumentation for
  Astronomy IV. p. 84461V, \mn@doi{10.1117/12.925738}

\bibitem[\protect\citeauthoryear{{Counselman}}{{Counselman}}{1973}]{counselman1973}
{Counselman} III C.~C.,  1973, \mn@doi [\apj] {10.1086/151964}, \href
  {http://adsabs.harvard.edu/abs/1973ApJ...180..307C} {180, 307}

\bibitem[\protect\citeauthoryear{Darwin}{Darwin}{1879}]{darwin1879}
Darwin G.~H.,  1879, \mn@doi [Proceedings of the Royal Society of London]
  {10.1098/rspl.1879.0028}, 29, 168

\bibitem[\protect\citeauthoryear{{Dawson}, {Murray-Clay}  \&
  {Johnson}}{{Dawson} et~al.}{2015}]{dawson2015}
{Dawson} R.~I.,  {Murray-Clay} R.~A.,   {Johnson} J.~A.,  2015, \mn@doi [\apj]
  {10.1088/0004-637X/798/2/66}, \href
  {http://adsabs.harvard.edu/abs/2015ApJ...798...66D} {798, 66}

\bibitem[\protect\citeauthoryear{{Doyle} et~al.,}{{Doyle}
  et~al.}{2013}]{doyle2013}
{Doyle} A.~P.,  et~al., 2013, \mn@doi [\mnras] {10.1093/mnras/sts267}, 428,
  3164

\bibitem[\protect\citeauthoryear{{Doyle}, {Davies}, {Smalley}, {Chaplin}  \&
  {Elsworth}}{{Doyle} et~al.}{2014}]{doyle2014}
{Doyle} A.~P.,  {Davies} G.~R.,  {Smalley} B.,  {Chaplin} W.~J.,   {Elsworth}
  Y.,  2014, \mn@doi [\mnras] {10.1093/mnras/stu1692}, 444, 3592

\bibitem[\protect\citeauthoryear{{Enoch}, {Collier Cameron}, {Parley}  \&
  {Hebb}}{{Enoch} et~al.}{2010}]{enoch2010}
{Enoch} B.,  {Collier Cameron} A.,  {Parley} N.~R.,   {Hebb} L.,  2010, \mn@doi
  [\aap] {10.1051/0004-6361/201014326}, \href
  {http://adsabs.harvard.edu/abs/2010A%26A...516A..33E} {516, A33}

\bibitem[\protect\citeauthoryear{{Fabrycky} \& {Tremaine}}{{Fabrycky} \&
  {Tremaine}}{2007}]{fabrycky2007}
{Fabrycky} D.,  {Tremaine} S.,  2007, \mn@doi [\apj] {10.1086/521702}, \href
  {http://adsabs.harvard.edu/abs/2007ApJ...669.1298F} {669, 1298}

\bibitem[\protect\citeauthoryear{{Ford} \& {Rasio}}{{Ford} \&
  {Rasio}}{2006}]{ford2006}
{Ford} E.~B.,  {Rasio} F.~A.,  2006, \mn@doi [\apjl] {10.1086/500734}, \href
  {http://adsabs.harvard.edu/abs/2006ApJ...638L..45F} {638, L45}

\bibitem[\protect\citeauthoryear{{Gibson} et~al.,}{{Gibson}
  et~al.}{2008}]{gibson2008}
{Gibson} N.~P.,  et~al., 2008, \mn@doi [\aap] {10.1051/0004-6361:200811015},
  \href {http://adsabs.harvard.edu/abs/2008A%26A...492..603G} {492, 603}

\bibitem[\protect\citeauthoryear{{Gray}}{{Gray}}{2008}]{gray2008}
{Gray} D.~F.,  2008, {The Observation and Analysis of Stellar Photospheres}.
Cambridge University Press, 2008

\bibitem[\protect\citeauthoryear{{Holmes} et~al.,}{{Holmes}
  et~al.}{2011}]{holmes2011}
{Holmes} S.,  et~al., 2011, \mn@doi [\pasp] {10.1086/662148}, \href
  {http://adsabs.harvard.edu/abs/2011PASP..123.1177H} {123, 1177}

\bibitem[\protect\citeauthoryear{{Howard} et~al.,}{{Howard}
  et~al.}{2012}]{howard2012}
{Howard} A.~W.,  et~al., 2012, \mn@doi [\apjs] {10.1088/0067-0049/201/2/15},
  \href {http://adsabs.harvard.edu/abs/2012ApJS..201...15H} {201, 15}

\bibitem[\protect\citeauthoryear{{Howell} et~al.,}{{Howell}
  et~al.}{2014}]{howell2014}
{Howell} S.~B.,  et~al., 2014, \mn@doi [\pasp] {10.1086/676406}, \href
  {http://adsabs.harvard.edu/abs/2014PASP..126..398H} {126, 398}

\bibitem[\protect\citeauthoryear{{Hut}}{{Hut}}{1980}]{hut1980}
{Hut} P.,  1980, \aap, \href
  {http://adsabs.harvard.edu/abs/1980A%26A....92..167H} {92, 167}

\bibitem[\protect\citeauthoryear{{Jehin} et~al.,}{{Jehin}
  et~al.}{2011}]{jehin2011}
{Jehin} E.,  et~al., 2011, The Messenger, \href
  {http://adsabs.harvard.edu/abs/2011Msngr.145....2J} {145, 2}

\bibitem[\protect\citeauthoryear{Kass \& Raftery}{Kass \&
  Raftery}{1995}]{kass1995}
Kass R.~E.,  Raftery A.~E.,  1995, Journal of the american statistical
  association, 90, 773

\bibitem[\protect\citeauthoryear{{Kolb}}{{Kolb}}{2014}]{kolb2014}
{Kolb} U.,  2014, in Revista Mexicana de Astronomia y Astrofisica Conference
  Series. pp 16--19

\bibitem[\protect\citeauthoryear{{Kov{\'a}cs}, {Zucker}  \&
  {Mazeh}}{{Kov{\'a}cs} et~al.}{2002}]{kovacs2002}
{Kov{\'a}cs} G.,  {Zucker} S.,   {Mazeh} T.,  2002, \mn@doi [\aap]
  {10.1051/0004-6361:20020802}, \href
  {http://adsabs.harvard.edu/abs/2002A%26A...391..369K} {391, 369}

\bibitem[\protect\citeauthoryear{{Lendl} et~al.,}{{Lendl}
  et~al.}{2012}]{lendl2012}
{Lendl} M.,  et~al., 2012, \mn@doi [\aap] {10.1051/0004-6361/201219585}, \href
  {http://adsabs.harvard.edu/abs/2012A%26A...544A..72L} {544, A72}

\bibitem[\protect\citeauthoryear{{Magain}}{{Magain}}{1984}]{magain1984}
{Magain} P.,  1984, \aap, \href
  {http://adsabs.harvard.edu/abs/1984A%26A...134..189M} {134, 189}

\bibitem[\protect\citeauthoryear{{Mandel} \& {Agol}}{{Mandel} \&
  {Agol}}{2002}]{mandel2002}
{Mandel} K.,  {Agol} E.,  2002, \mn@doi [\apjl] {10.1086/345520}, \href
  {http://adsabs.harvard.edu/abs/2002ApJ...580L.171M} {580, L171}

\bibitem[\protect\citeauthoryear{{Matsumura}, {Peale}  \& {Rasio}}{{Matsumura}
  et~al.}{2010}]{matsumura2010}
{Matsumura} S.,  {Peale} S.~J.,   {Rasio} F.~A.,  2010, \mn@doi [\apj]
  {10.1088/0004-637X/725/2/1995}, \href
  {http://adsabs.harvard.edu/abs/2010ApJ...725.1995M} {725, 1995}

\bibitem[\protect\citeauthoryear{{Maxted} et~al.,}{{Maxted}
  et~al.}{2011}]{maxted2011}
{Maxted} P.~F.~L.,  et~al., 2011, \mn@doi [\pasp] {10.1086/660007}, 123, 547

\bibitem[\protect\citeauthoryear{{Maxted}, {Serenelli}  \&
  {Southworth}}{{Maxted} et~al.}{2015}]{maxted2015}
{Maxted} P.~F.~L.,  {Serenelli} A.~M.,   {Southworth} J.,  2015, \mn@doi [\aap]
  {10.1051/0004-6361/201425331}, \href
  {http://adsabs.harvard.edu/abs/2015A%26A...575A..36M} {575, A36}

\bibitem[\protect\citeauthoryear{{McCormac}, {Skillen}, {Pollacco}, {Faedi},
  {Ramsay}, {Dhillon}, {Todd}  \& {Gonzalez}}{{McCormac}
  et~al.}{2014}]{mccormac2014}
{McCormac} J.,  {Skillen} I.,  {Pollacco} D.,  {Faedi} F.,  {Ramsay} G.,
  {Dhillon} V.~S.,  {Todd} I.,   {Gonzalez} A.,  2014, \mn@doi [\mnras]
  {10.1093/mnras/stt2449}, 438, 3383

\bibitem[\protect\citeauthoryear{{McLaughlin}}{{McLaughlin}}{1924}]{mclaughlin1924}
{McLaughlin} D.~B.,  1924, \mn@doi [\apj] {10.1086/142826}, \href
  {http://adsabs.harvard.edu/abs/1924ApJ....60...22M} {60}

\bibitem[\protect\citeauthoryear{{Mottram}, {Steele}  \& {Morales}}{{Mottram}
  et~al.}{2004}]{mottram2004}
{Mottram} C.~J.,  {Steele} I.~A.,   {Morales} L.,  2004, in {Moorwood}
  A.~F.~M.,  {Iye} M.,  eds,  \procspie Vol. 5492, Ground-based Instrumentation
  for Astronomy. pp 677--688, \mn@doi{10.1117/12.551337}

\bibitem[\protect\citeauthoryear{{Munari} \& {Zwitter}}{{Munari} \&
  {Zwitter}}{1997}]{munari1997}
{Munari} U.,  {Zwitter} T.,  1997, \aap, \href
  {http://adsabs.harvard.edu/abs/1997A%26A...318..269M} {318, 269}

\bibitem[\protect\citeauthoryear{{Mustill}, {Davies}  \& {Johansen}}{{Mustill}
  et~al.}{2015}]{mustill2015}
{Mustill} A.~J.,  {Davies} M.~B.,   {Johansen} A.,  2015, \mn@doi [\apj]
  {10.1088/0004-637X/808/1/14}, \href
  {http://adsabs.harvard.edu/abs/2015ApJ...808...14M} {808, 14}

\bibitem[\protect\citeauthoryear{{Neveu-VanMalle} et~al.,}{{Neveu-VanMalle}
  et~al.}{2016}]{neveuvanmalle2015}
{Neveu-VanMalle} M.,  et~al., 2016, \mn@doi [\aap]
  {10.1051/0004-6361/201526965}, \href
  {http://adsabs.harvard.edu/abs/2016A%26A...586A..93N} {586, A93}

\bibitem[\protect\citeauthoryear{{Ohta}, {Taruya}  \& {Suto}}{{Ohta}
  et~al.}{2005}]{ohta2005}
{Ohta} Y.,  {Taruya} A.,   {Suto} Y.,  2005, \mn@doi [\apj] {10.1086/428344},
  \href {http://adsabs.harvard.edu/abs/2005ApJ...622.1118O} {622, 1118}

\bibitem[\protect\citeauthoryear{{Penev} \& {Sasselov}}{{Penev} \&
  {Sasselov}}{2011}]{penev2011}
{Penev} K.,  {Sasselov} D.,  2011, \mn@doi [\apj] {10.1088/0004-637X/731/1/67},
  \href {http://adsabs.harvard.edu/abs/2011ApJ...731...67P} {731, 67}

\bibitem[\protect\citeauthoryear{{Pepe}, {Mayor}, {Galland}, {Naef}, {Queloz},
  {Santos}, {Udry}  \& {Burnet}}{{Pepe} et~al.}{2002}]{pepe2002}
{Pepe} F.,  {Mayor} M.,  {Galland} F.,  {Naef} D.,  {Queloz} D.,  {Santos}
  N.~C.,  {Udry} S.,   {Burnet} M.,  2002, \mn@doi [\aap]
  {10.1051/0004-6361:20020433}, \href
  {http://adsabs.harvard.edu/abs/2002A%26A...388..632P} {388, 632}

\bibitem[\protect\citeauthoryear{{Perruchot} et~al.,}{{Perruchot}
  et~al.}{2008}]{perruchot2008}
{Perruchot} S.,  et~al., 2008, in Ground-based and Airborne Instrumentation for
  Astronomy II. p. 70140J, \mn@doi{10.1117/12.787379}

\bibitem[\protect\citeauthoryear{{Pollacco} et~al.,}{{Pollacco}
  et~al.}{2006}]{pollacco2006}
{Pollacco} D.~L.,  et~al., 2006, \mn@doi [\pasp] {10.1086/508556}, \href
  {http://adsabs.harvard.edu/abs/2006PASP..118.1407P} {118, 1407}

\bibitem[\protect\citeauthoryear{{Pollacco} et~al.,}{{Pollacco}
  et~al.}{2008}]{pollacco2008}
{Pollacco} D.,  et~al., 2008, \mn@doi [\mnras]
  {10.1111/j.1365-2966.2008.12939.x}, \href
  {http://adsabs.harvard.edu/abs/2008MNRAS.385.1576P} {385, 1576}

\bibitem[\protect\citeauthoryear{{Queloz} et~al.,}{{Queloz}
  et~al.}{2000}]{queloz2000}
{Queloz} D.,  et~al., 2000, \aap, \href
  {http://adsabs.harvard.edu/abs/2000A%26A...354...99Q} {354, 99}

\bibitem[\protect\citeauthoryear{{Rasio} \& {Ford}}{{Rasio} \&
  {Ford}}{1996}]{rasio1996}
{Rasio} F.~A.,  {Ford} E.~B.,  1996, \mn@doi [Science]
  {10.1126/science.274.5289.954}, \href
  {http://adsabs.harvard.edu/abs/1996Sci...274..954R} {274, 954}

\bibitem[\protect\citeauthoryear{{Rossiter}}{{Rossiter}}{1924}]{rossiter1924}
{Rossiter} R.~A.,  1924, \mn@doi [\apj] {10.1086/142825}, \href
  {http://adsabs.harvard.edu/abs/1924ApJ....60...15R} {60}

\bibitem[\protect\citeauthoryear{{Santerne} et~al.,}{{Santerne}
  et~al.}{2015}]{santerne2015}
{Santerne} A.,  et~al., 2015, \mn@doi [\mnras] {10.1093/mnras/stv1080}, \href
  {http://adsabs.harvard.edu/abs/2015MNRAS.451.2337S} {451, 2337}

\bibitem[\protect\citeauthoryear{Schwarz et~al.}{Schwarz
  et~al.}{1978}]{schwarz1978}
Schwarz G.,  et~al., 1978, The annals of statistics, 6, 461

\bibitem[\protect\citeauthoryear{{Sestito} \& {Randich}}{{Sestito} \&
  {Randich}}{2005}]{sestito2005}
{Sestito} P.,  {Randich} S.,  2005, \mn@doi [\aap]
  {10.1051/0004-6361:20053482}, \href
  {http://adsabs.harvard.edu/abs/2005A%26A...442..615S} {442, 615}

\bibitem[\protect\citeauthoryear{{Sing} et~al.,}{{Sing}
  et~al.}{2016}]{sing2016}
{Sing} D.~K.,  et~al., 2016, \mn@doi [\nat] {10.1038/nature16068}, \href
  {http://adsabs.harvard.edu/abs/2016Natur.529...59S} {529, 59}

\bibitem[\protect\citeauthoryear{{Steele} et~al.,}{{Steele}
  et~al.}{2004}]{steele2004}
{Steele} I.~A.,  et~al., 2004, in {Oschmann} Jr. J.~M.,  ed.,  \procspie Vol.
  5489, Ground-based Telescopes. pp 679--692, \mn@doi{10.1117/12.551456}

\bibitem[\protect\citeauthoryear{{Steele}, {Bates}, {Gibson}, {Keenan},
  {Meaburn}, {Mottram}, {Pollacco}  \& {Todd}}{{Steele}
  et~al.}{2008}]{steele2008}
{Steele} I.~A.,  {Bates} S.~D.,  {Gibson} N.,  {Keenan} F.,  {Meaburn} J.,
  {Mottram} C.~J.,  {Pollacco} D.,   {Todd} I.,  2008, in Ground-based and
  Airborne Instrumentation for Astronomy II. p. 70146J (\mn@eprint {arXiv}
  {0809.3351}), \mn@doi{10.1117/12.787889}

\bibitem[\protect\citeauthoryear{{Steffen} et~al.,}{{Steffen}
  et~al.}{2012}]{steffen2012}
{Steffen} J.~H.,  et~al., 2012, \mn@doi [Proceedings of the National Academy of
  Science] {10.1073/pnas.1120970109}, \href
  {http://adsabs.harvard.edu/abs/2012PNAS..109.7982S} {109, 7982}

\bibitem[\protect\citeauthoryear{{Stetson}}{{Stetson}}{1987}]{stetson1987}
{Stetson} P.~B.,  1987, \mn@doi [\pasp] {10.1086/131977}, \href
  {http://adsabs.harvard.edu/abs/1987PASP...99..191S} {99, 191}

\bibitem[\protect\citeauthoryear{{Torres}, {Andersen}  \&
  {Gim{\'e}nez}}{{Torres} et~al.}{2010}]{torres2010}
{Torres} G.,  {Andersen} J.,   {Gim{\'e}nez} A.,  2010, \mn@doi [\aapr]
  {10.1007/s00159-009-0025-1}, \href
  {http://adsabs.harvard.edu/abs/2010A%26ARv..18...67T} {18, 67}

\bibitem[\protect\citeauthoryear{{Wang}, {Fischer}, {Horch}  \& {Huang}}{{Wang}
  et~al.}{2015}]{wang2015}
{Wang} J.,  {Fischer} D.~A.,  {Horch} E.~P.,   {Huang} X.,  2015, \mn@doi
  [\apj] {10.1088/0004-637X/799/2/229}, \href
  {http://adsabs.harvard.edu/abs/2015ApJ...799..229W} {799, 229}

\bibitem[\protect\citeauthoryear{{Weiss} \& {Schlattl}}{{Weiss} \&
  {Schlattl}}{2008}]{weiss2008}
{Weiss} A.,  {Schlattl} H.,  2008, \mn@doi [\apss] {10.1007/s10509-007-9606-5},
  \href {http://adsabs.harvard.edu/abs/2008Ap%26SS.316...99W} {316, 99}

\bibitem[\protect\citeauthoryear{{Winn}, {Fabrycky}, {Albrecht}  \&
  {Johnson}}{{Winn} et~al.}{2010}]{winn2010}
{Winn} J.~N.,  {Fabrycky} D.,  {Albrecht} S.,   {Johnson} J.~A.,  2010, \mn@doi
  [\apjl] {10.1088/2041-8205/718/2/L145}, \href
  {http://adsabs.harvard.edu/abs/2010ApJ...718L.145W} {718, L145}

\bibitem[\protect\citeauthoryear{{Wright}, {Marcy}, {Howard}, {Johnson},
  {Morton}  \& {Fischer}}{{Wright} et~al.}{2012}]{wright2012}
{Wright} J.~T.,  {Marcy} G.~W.,  {Howard} A.~W.,  {Johnson} J.~A.,  {Morton}
  T.~D.,   {Fischer} D.~A.,  2012, \mn@doi [\apj]
  {10.1088/0004-637X/753/2/160}, \href
  {http://adsabs.harvard.edu/abs/2012ApJ...753..160W} {753, 160}

\makeatother
\end{thebibliography}

\bsp

\label{lastpage}

\end{document}